\documentclass[draft]{agujournal2019}
\usepackage{url} 
\usepackage{lineno}
\usepackage{soul}
\usepackage{float}
\restylefloat{table}
\restylefloat{figure}

\usepackage{amsmath}
\usepackage{xcolor,colortbl} 
\usepackage{breakurl}        

\usepackage{multirow}
\usepackage[normalem]{ulem}
\renewcommand{\arraystretch}{2.4}
\definecolor{Gray}{gray}{0.95}
\newcolumntype{Z}{>{\columncolor{Gray}}c}
\definecolor{Grayb}{gray}{0.65}
\newcolumntype{G}{>{\columncolor{Grayb}}c}

\DeclareMathOperator*{\argmax}{argmax} 

\draftfalse

\journalname{JGR: Atmospheres}

\begin{document}

%
%

\title{Data Fusion of Total Solar Irradiance Composite Time Series Using 41 years of Satellite Measurements}

%
%




\authors{J.-P. Montillet\affil{1}, W. Finsterle\affil{1}, G. Kermarrec\affil{2}, R. Sikonja\affil{3}, M. Haberreiter\affil{1},W. Schmutz\affil{1}, T. Dudok de Wit\affil{4}}

\affiliation{1}{Physikalisch-Meteorologisches Observatorium Davos/World Radiation Center (PMOD/WRC), Davos, Switzerland}
\affiliation{2}{\textcolor{black}{Institute of meteorology and climatology, Leibniz University Hannover, Germany}}
\affiliation{3}{Department of Computer Science, Eidgenössische Technische Hochschule (ETH), Zurich, Switzerland}
\affiliation{4}{Laboratoire de Physique et Chimie de l'Environnement et de l'Espace, CNRS, CNES and University of Orléans, Orléans, France}





\correspondingauthor{J.-P. Montillet}{jean-philippe.montillet@pmodwrc.ch}




\begin{keypoints}


\item {Application of data fusion to merge $41$ years of satellite observations into a composite TSI time series}

\item  A comprehensive time-frequency analysis to characterise the solar cycle and the stochastic noise
\item Full investigation of variations at solar minima to distinguish between stochastic noise and possible underlying phenomena linked to the solar activity

\end{keypoints}

%
%

%
%


\begin{abstract}
Since the late $1970$'s, successive satellite missions have been monitoring the sun’s activity and recording the total solar irradiance (TSI). \textcolor{black}{Some of these measurements have lasted for more than a decade. In order to obtain a seamless record whose duration exceeds that of the individual instruments, the time series have to be merged.}
%
Climate models  can be better validated using such long TSI time series which can also 
help to provide stronger constraints on past climate reconstructions (e.g.,back to the Maunder minimum). 
We propose a $3$-step method based on data fusion, including a stochastic noise model to take into account short and long-term correlations. Compared with previous products scaled at the nominal TSI value of $\sim 1361$ W/m$^2$, \textcolor{black}{the difference  is below $0.2$ W/m$^2$ in terms of solar minima.}
Next, we model the frequency spectrum of this $41$-year TSI composite time series with a Generalized  Gauss-Markov model to help describe an observed flattening at high frequencies. It allows us to fit a linear trend into these TSI time series by joint inversion with the stochastic noise model via a maximum-likelihood estimator. \textcolor{black}{Our results show that the amplitude of such trend is $ \sim -0.004$ $\pm$ $0.004$  W/(m$^2 yr$) for the period $1980$-$2021$.} These results are compared with the difference of irradiance values estimated from two consecutive solar minima. We conclude that the trend in these composite time series is mostly an artifact due to the coloured noise.
\end{abstract}



%
%

\section{Introduction}
     \label{S-Introduction} 


Monitoring the Earth's radiation budget is a key aspect in understanding the anthropogenic contribution to climate forcing \cite{Kren2015}. 
Total solar irradiance is Earth's dominant energy input. Global temperature and TSI are linked by the energy equilibrium equation for the Earth system. As summarized by \citeA{Werner2021}, the derivation of this equation with respect to a variation of the solar irradiance has two terms: a direct forcing term, which can be derived analytically and quantified accurately from the Stefan-Boltzmann law, and a second term, describing indirect influences on the surface temperature. \textcolor{black}{If a small TSI variation should force a large temperature variation, then it has to be the second indirect term that strongly amplifies the effect of the direct forcing.} This amplification mechanism has been debated in the scientific community for the past two decades \cite{DavidH, Shapiro2017,Egorova, Werner2021}, because it will most likely call for a strong modification of the models that describe the Earth’s climate response to variations in the solar radiative output.
\textcolor{black}{On shorter time scales (e.g., weekly), the existence of trend in the measurements could on a longer timescale (e.g., yearly) significantly bias the analysis of a solar phenomena (e.g., estimation of a new solar minima).}
Therefore, it is important to produce robust and reliable TSI composite time series using all the observations available recorded by successive space instruments spanning $4$ decades. Satellite measurements show that TSI varies on all timescales with a pronounced quasi-periodicity signature of approximately $11$ years \cite{Froehlich1997b,Kopp2016}. Timescale variations can be classified in subdaily (minutes to hour), daily to weekly, and yearly to one solar cycle. Major mechanisms, such as the evolution of magnetic features on the solar surface, which dominate each timescale are complex and still under investigation within the solar physics community \cite{Yeo2017SolarIV,XiangB2019}. 
 Several studies \cite{Fontenla2009,Kopp2011,Yeo2021} have shown that TSI variations on timescales of hours are a combination of sunspot blocking and an intensification due to bright faculae, plages and other elements. This makes forecasting and modelling the solar cycle more difficult. As all satellite observations are limited in time, constructing composites is a key aspect to the investigation of TSI over several decades. Merging all these observations is a difficult exercise with both a scientiﬁc and a statistical challenge \cite{Dudok2017}. Previous approaches \cite{Willson1997,Frohlich2004,Mekaoui2008} produced TSI composite time series by daisy chaining all the available TSI observations, but without including any models of the stochastic noise properties. The first methodology which relied on some knowledge of the underlining noise characteristics was developed by \citeA{Dudok2017}, including a data-driven noise model and a multiscale decomposition. 
{ An approach along these lines was also employed by \cite{Schoell2016} and \cite{Haberreiter2017}. The authors made use of all available solar spectral irradiance datasets.}

    Here, we present a different statistical approach, which is based on three key steps. The first step relies on data fusion of multiple observations based on a Bayesian framework and Gaussian processes. Our composite spanning the last $4$ decades is obtained in the second step by daisy chaining the sub-time series resulting from the first step. 
    The last step is the application of wavelet filtering to correct some unwanted correlations in the fused observations (i.e. bandwidth noise). The robustness of our approach is guaranteed via careful modelling of the TSI observations during the data fusion process. Various assumptions formulated by data scientists, can introduce biases in the data analysis. Some algorithms \cite{ Willson1997,Frohlich2004,Mekaoui2008} based on daisy chaining the raw TSI observations required the choice of the most trustworthy instrument, hence introducing a bias toward preconceived ideas of how the TSI should vary. Note that our data fusion process merges datasets from subsequent solar missions based on a few stochastic noise assumptions.
    It circumvents the weakness of choosing the most trustworthy instrument when performing the daisy chain on the TSI observations from various instruments, which could influence towards preconceived ideas of how the TSI should vary \cite{Dudok2017}.
    
    
    {The motivation to develop \textcolor{black}{our new} methodology is to advance the data-driven approach first adopted by \citeA{Dudok2017}. \textcolor{black}{Our} resulting TSI composite corresponds with the instrument-driven approach by the late Dr. Claus Fröhlich \cite{Froehlich2006} when calibrated in a similar way. We use his composite \cite{Froehlich2006} as a baseline for other aspects as well (i.e. Step $3$ of the algorithm). \textcolor{black}{However,} the TSI community consensus composite \textcolor{black}{developed by \citeA{Dudok2017}}, is the reference time series against which our new results are compared.}
    
    \vspace{0.5em}
    
    Finally, recent studies \cite{Scafetta2020,Dudok2020,Werner2021} have debated about the existence of a trend in the TSI composite time series. If it exists, the origin of this trend in the TSI observations is unknown: one could speculate that it could be caused by a drift in the peak amplitude of solar cycles while the minima could all remain at the same level. Another possibility is the presence of an unknown diffusion process which could generate a transient signal making variations in the solar minima. {One possibility could be that the brightness of the quiet Sun shows a trend, as suggested by \citeA{Shapiro2011}}. \citeA{Dudok2020} argue in favor of an artifact generated by unwanted noise looking at the difference between consecutive solar minima. Here, we go further by performing a time-frequency analysis of various TSI composite time series produced with various techniques, including our new product. We focus on describing the stochastic noise properties within these $40$-year long time series. We use this knowledge to model the TSI composites and to conclude on the existence of a long-term trend.
\section{Description of The Raw Datasets} \label{NewSection2} 

Table \ref{Table1} displays the instruments and the processing centers providing the observations relative to the various missions used in this study. The data processing, including corrections for all a priori known influences such as the distance from the sun (normalized to $1$ AU), radial velocity of the sun, and thermal, optical, and electrical corrections, are usually implemented by each processing center, leading to level-$1$ time series. Most of these instruments observe on a daily basis, with occasional interruptions and outliers. Usually, one to three of them operate simultaneously, although some days are devoid of observations. Note that {\it PMODv21a} is the new VIRGO/SOHO dataset released in  March $2021$ by PMOD using a new software described in \citeA{Finsterle2021}.  
{\it PREMOS (v1)}  is the released version described in 
\citeA{Schmutz_etal2013}. ERBE/ERBS and  HF/NIMBUS-7 ERB datasets are retrieved from the PMOD archive and the corrections  made by C. Fröhlich, which are explained in \citeA{Frohlich2006}.

\begin{table}[htb!]
\scriptsize
\begin{center}
\caption{
Overview of the datasets used in this study including the start and end dates for each mission and the latest version  released by the various centers. 
} 
 \begin{tabular}{||c| c | c c||}
 \hline
 Mission/Experiment/Instrument & Version & Start Date & End Date \\ 
 \hline
 HF/NIMBUS‐7 ERB  & -  &  11/1978 & 1/1993 \\
 ERBE/ERBS 	& - &	10/1984 &	8/2003   \\
 \hline
 VIRGO/SOHO  & PMODv21a & 01/1996 & active \\ 
 PREMOS/PICARD  & v1 & 06/2010 & 03/2014 \\
 \hline
 ACRIM1/SMM & 1 & 2/1980 &	7/1989  \\ 
 ACRIM2/UARS & 7/14 & 10/1991 & 9/2000  \\ 
 ACRIM3/ACRIMSAT  &  11/13 & 04/2000 & 11/2013 \\ 
 TIM/SORCE & 19 & 02/2003 & 02/2020 \\
 TIM/TCTE & {4} & 12/2013 & 05/2019 \\
 TIM/TSIS   & 3 & 11/01/2018 & active \\
  \hline
\end{tabular} \label{Table1}
\end{center}
\end{table}

Figure \ref{FigureX1} displays the observations from each mission spanning a specific period of time. All space missions have provided TSI observations with a different sampling rate. Recent instruments make several observations per day (with a cadence of up to $50$ s for TIM/SORCE). Earlier radiometers such as ERBE/ERBS observes the sun once every $14$ days for $3$ min on average, so that the stochastic noise properties of such sensors are different to those with a higher recording rate. Note that {\it active} in Table  \ref{Table1} means that the instrument is still operating. The dataset from these missions ends in March $2021$ for this study.
    
\section{The 3-step Method to Produce the $41$-year Long TSI Composite}
     \label{Section2} 

\subsection{Step $1$: Merging Multiple Datasets with Data Fusion}
\label{Section2.1}
Data fusion is the process of integrating multiple data sources to produce more consistent, accurate, and useful information than that provided by each individual data source alone. The process has found many applications in various areas ranging from industry to geosciences and solar science \cite{Cocchi2019}.

\begin{figure}[htbp!]
\hspace{-5em}
\includegraphics[width=1.4\textwidth]{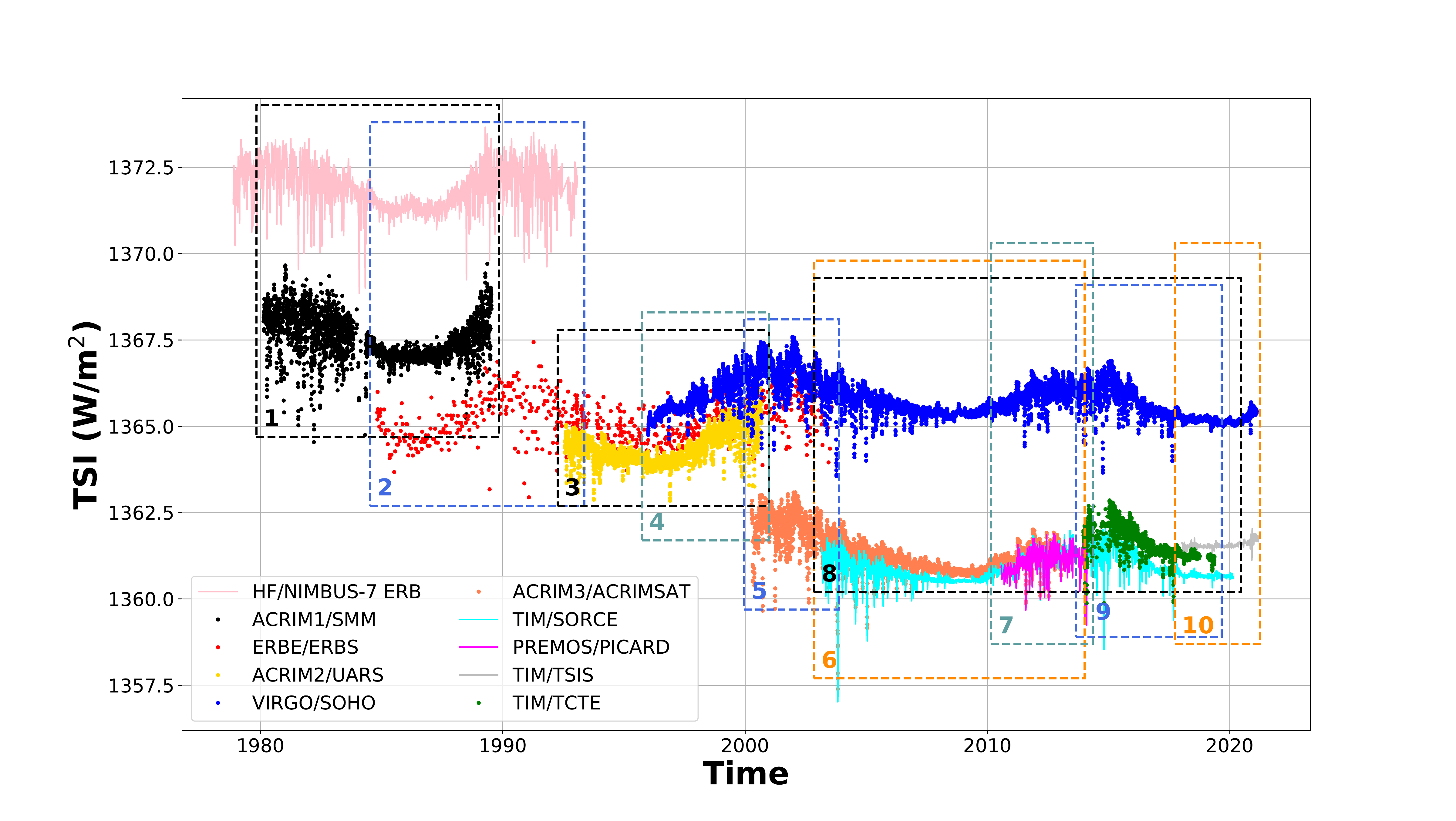}
\caption{Various satellite missions which have conduscted TSI observations since the late $1970$'s. We perform the fusion using \textcolor{black}{the observations included in  each box} (dash lines).}\label{FigureX1}
\end{figure}

Let us call the observations to merge $a(t_i), b(t_i), c(t_i)$ (with ${\{i=[1,n]\}}$), which are recorded using three different instruments. \textcolor{black}{The noise for each observation is additive and uncorrelated between the instruments. The model of the observations is defined such as}:

\begin{equation}\label{CorrectedMeas}
\left\{ 
  \begin{array}{l l}
    \hspace{-0.3em} a(t_i)   = s(t_i)  + \epsilon_a(t_i), & \hspace{0.5em}   \epsilon_a(t_i)  \sim \mathcal{N}(0, \sigma^2_a) \\
     \hspace{-0.3em} b(t_i)  = s(t_i)  + \epsilon_b(t_i), & \hspace{0.5em}   \epsilon_b(t_i)   \sim \mathcal{N}(0, \sigma^2_b) \\
     \hspace{-0.3em} c(t_i)  = s(t_i)  + \epsilon_c(t_i), & \hspace{0.5em}   \epsilon_c(t_i)  \sim \mathcal{N}(0, \sigma^2_c) \\
  \end{array} \right.
\end{equation}

where $\epsilon_a$, $\epsilon_b$ and $\epsilon_c$  are zero-mean Gaussian distributed random variables (with variance $\sigma_a^2$, $\sigma_b^2$ and $\sigma_c^2$ respectively) modelling the noise properties intrinsic to each instrument. The data fusion algorithm aims to merge the observations available at each epoch $t_i$ in order to get a reliable estimate of the true signal $s$, i.e. the solar activity \cite{Feynman1982}. We formulate the following assumptions: i) the solar cycle is an unknown process (i.e.  not a perfect sinusoidal signal with a $11.5$ year cycle) and its variations are random (no a priori knowledge). Physically, it means that two or more radiometers monitor the solar activity from a different distance due to different orbits, but monitoring the same underlying information on the solar cycle. The model of $s$ is a Gaussian process (GP) with zero mean and a covariance function $k_\mathbf{\theta}$ (or kernel). A GP can be generally defined as a finite sum of random variables normally distributed where the overall distribution is a multivariate normal distribution \cite{Kolar2020}; ii) we consider the noise on the measurements zero-mean Gaussian distributed. We can then estimate the parameters of the  model of $s(t)$ via a maximum likelihood estimator (MLE). Therefore, we have $\mathbf{s} \sim$ $GP (0, k_{\mathbf{\theta}}(t_i, t_i)_{\{i=[1,n]\}})$, with $n$ the number of samples in the various measurements $a$, $b$ and $c$. The parameters of $s(t)$, expressed in $\mathbf{\theta}$, are selected by maximizing the {log-marginal likelihood} $ \log{p(\mathbf{y}|\mathbf{x})}$, where  $\mathbf{y}$ and $\mathbf{x}$ are \textcolor{black}{the corrected observations, i.e. $\mathbf{y} = [a(t_i), b(t_i), c(t_i)]$, and the corresponding time, i.e. $\mathbf{x} = [t_i, t_i]$.} \textcolor{black}{The main limitation of GPs is that given $n$ observations, the inverse of the $n$-by-$n$ covariance matrix must be computed. Time complexity of such operation is of the order of $O(n^3)$, which is computationally expensive for long records. Some of these missions have been recording data over two decades, which has generated large datasets. To overcome this limitation, we approximate the exact GPs by utilizing sparse Gaussian processes (SGP),} yielding a maximization problem of the lower bound of $\log{p(\mathbf{y}|\mathbf{x})}$ following \citeA{Bauer2016}:
\begin{equation}\label{lowerboundP}
    \log{p(\mathbf{y}|\mathbf{x})} \geq - \frac{1}{2} \mathbf{y}^T(\mathbf{Q}_{\mathbf{\theta}}+ \sigma^2 \mathbf{I})^{-1} \mathbf{y}  - \frac{1}{2} \log{| \mathbf{Q}_{\mathbf{\theta}} + \sigma^2 \mathbf{I}|} - \frac{n}{2} \log{(2 \pi} ) - \frac{1}{2\sigma^2} tr( k_{\mathbf{\theta}} (\mathbf{x},\mathbf{x}) - \mathbf{Q}_{\mathbf{\theta}})
\end{equation}
where $tr$ is the trace operator, $Q_{\mathbf{\theta}}  = k_{\mathbf{\theta}} (\mathbf{x},\mathbf{u}) k_{\mathbf{\theta}} (\mathbf{u},\mathbf{u}) ^{-1} k_{\mathbf{\theta}} (\mathbf{u},\mathbf{x})$,  $\mathbf{u}$ is a vector of inducing points to learn about the stochastic properties of the data, which allows to take into account long-term and short-term correlations in the observations and are a reasonable approximation of $s$. This subset of observations is used to estimate the initial parameters in $\mathbf{\theta}$. $\mathbf{I}$ is the identity matrix, with $\sigma^2 \mathbf{I}$ the noise component of the covariance matrix (assuming uncorrelated measurements) formulated as $diag([\sigma^2_a, \sigma^2_b,$  $\sigma^2_c, \sigma^2_a, \sigma^2_b, \sigma^2_c, ...])$. Next, we estimate the kernel $k_{\mathbf{\theta}}$ by maximizing the right-hand-side of Eq. \eqref{lowerboundP} with respect to  $\mathbf{u}$ and $\mathbf{\theta}$. {Further mathematical simplifications to estimate the kernel are voluntarily left out for clarity, but readers can refer to \citeA{Kolar2020}. We also define comprehensively what is a GP and its application to TSI time series (including the Bayesian framework) in \ref{D-appendix}.} We emphasize that
the number of inducing points defines the size of the matrix $\mathbf{Q}_{\mathbf{\theta}}$ which must be inverted in the maximization  of Eq. \eqref{lowerboundP}. \textcolor{black}{A large number of points is necessary to avoid completely smoothing the short-term and long-term correlations due to the difference in the recording rate of the instruments.} The computational complexity is on the order of $O(nm^2)$ (with $m$ the number of inducing points, $m$ $\ll$ $n$ ).  Therefore, we are limited by the computing resources available when dealing with a large matrix (i.e. over $m=3000$). 
Now, the number of inducing points vary due to the size of the input datasets (i.e. size of the boxes defined in Figure \ref{FigureX1}). \textcolor{black}{ \ref{C-appendix} further documents the influence of this parameter on the quality of the fused time series.}  
We chose $2000$ points, which provides a good balance between computational time and accuracy. Now, most of the sub-time series have a length greater than $7$ years, therefore one can select $2500$ points or more if necessary. The shortest time series is when fusing PREMOS/PICARD, VIRGO/SOHO and TIM/SORCE (box $7$), where we have  $\sim$ $1400 $ observations. In this case, we use $1300$ points. Note that the number of inducing points for the fusion of TIM/TSIS and VIRGO/SOHO is also constrained for the same reasons.

   As described in Section \ref{NewSection2}, each instrument records the data with a different sampling rate. The fusion requires regularly sampled records with no gaps. \textcolor{black}{We ﬁrst regrid all the datasets with a sampling rate of $1$ day}. The datasets recorded with a lower rate are (linearly) interpolated. Note that the starting date of the composite time series is defined by the fusion between {HF/NIMBUS-7 ERB}  and {ACRIM1/SMM} which is February $1980$.
 
 Nonetheless, events resulting from short-term variations in solar activity lasting less than a few days are relatively difficult to fuse. The radiometers on board of the various missions at time $t_i$ may not have recorded exactly the same event due to different distances (i.e. different orbits) and also because of differences in the observation time (i.e. sampling rate). \textcolor{black}{The fusion of these short-term solar variations generally results in keeping only major events or underlying long-term solar events recorded by all the instruments at a specific time ($t_i$)}.
 
\textcolor{black}{Last, we should distinguish in the following between stochastic and solar noise in order to avoid any confusion. We refer to stochastic noise as the statistical definition of random processes which includes short and long-term correlations} (i.e. white and coloured noise). Solar noise results from photospheric activity associated with granules varying at different timescales over a few hours (e.g., sunspots) to a decade (e.g., solar cycle), which generate fluctuations in the recorded irradiance values. Further discussions are included in \ref{D-appendix}.

\subsection{Step $2$: Producing the $41$-year composite time series with a modified adaptive filter}
\label{Section2.2}

 To perform the data fusion, we first select all the periods where at least two missions overlap for more than $6$ months (see boxes in Figure \ref{FigureX1}). With a shorter overlapping time, simulations have shown that the fusion is not optimal due to the limited number of inducing points. 
 \textcolor{black}{For each overlapping period, we fuse the time series corresponding to different missions/instruments together in order to obtain  the sub-time series.}

    We produce $q$ partially overlapping composite time series ($y_q$) with associated uncertainties ($ \alpha^2_q$). We use a modified adaptive algorithm  \cite{Haykin2004} to daisy-chain all the sub-time series  and build the $41$-year composite as follows:
%
%

\begin{equation}\label{DaisyChainAlgo}
\left\{ 
  \begin{array}{l l}
    \hspace{-0.3em} y(t_i)   =&  y_1(t_i) w(t_i)  + y_2(t_i) (1-w(t_i)) \\
     \hspace{-0.3em} w(t_i)  = &    \alpha^2_1(t_i) / (\alpha^2_2 (t_i) + \alpha^2_1 (t_i) ) \\
     \hspace{-0.3em} \bar{\alpha^2_1} \leq \bar{\alpha^2_2}\\
     \hspace{-0.3em} \alpha^2(t_i) = &    0.5 (\alpha^2_2 (t_i) + \alpha^2_1 (t_i) )  \\
  \end{array} \right.
\end{equation}

with $t_i$ the time spanning the period 1978-2021 and daily sampling. The two overlapping time series are $y_1$ and $y_2$ and associated uncertainties $\alpha^2_1$ and $\alpha^2_2$ respectively. $\bar{\alpha^2_1}$, $\bar{\alpha^2_2}$ are the average of the uncertainties over the overlapping time for $y_1$ and $y_2$. Note that $y_1$ is chosen in order to satisfy the condition $\bar{\alpha^2_1} \leq \bar{\alpha^2_2}$. \textcolor{black}{This condition is necessary to guarantee that $w$ is in the interval $[0,1]$. We define $w$ using the denominator $\alpha^2_2 (t_i) + \alpha^2_1 (t_i) $ in order to avoid any divergence. We exclude the case for which $\alpha^2_1 (t_i)$ $=\alpha^2_2 (t_i)$ $=0$.}

    The mean value  of each sub-time series resulting from the data fusion process is relative to the lowest mean value of the input TSI datasets. We end up with a different mean value for each sub-time series. Before applying the modified adaptive algorithm on two consecutive time series, we scale the second sub-time series using the common period between the two time series. It results in a TSI composite time series with an arbitrary mean value. \textcolor{black}{To obtain the correctly scaled TSI composite, we employ the TSI value defined by \citeA{Prsa_etal2016}, which was derived as the averaged TSI value over Solar Cycle $23$}. This approach is also applied here, i.e. we determine the average TSI for Solar Cycle $23$ of the new composite and scale it to the nominal TSI value. As such the new TSI composite is consistent with the nominal TSI of $1361$ W/m$^2$ as recommended by the IAU 2015 Resolution B3. Finally, the last step includes a wavelet filter in order to smooth the correlations introduced by the data fusion. The effect of these correlations in time and frequency domains is discussed in the next section. 
\subsection{Step $3$: Filtering the Composite with a Wavelet Filter}
\label{Section2.3}
%
 \textcolor{black}{The data fusion process results in a filtering of the high frequency area of the spectrum of the composite time series. This is similar to the effect of a low pass filter, which would let the low frequencies pass through. Possible reasons are to be associated with the number of inducing points or other parameters as discussed in Section \ref{Section2.1}. This drawback is unwanted: 1) it may mask some peaks at high frequencies linked with daily components; 2) the solar minimum can be affected in the time domain by the presence of long-term correlations. Unfortunately, this effect is theoretically unpredictable. \textcolor{black}{Here we propose to make use of a wavelet filter to generate our final product. We intend to reconstruct empirically the high frequencies of our fused composite time series with respect to a reference one. We have chosen the one released by \citeA{Froehlich2006}, based on its properties in the time domain, discussed in the next section. To that end, the wavelet variance (WV) provides a robust mathematical framework to perform the rescaling of the high frequency noise of the composite time series \cite{Abry1998}.} More specifically, we decompose the time series into an ensemble of records whose spectral content is concentrated in a specific frequency band. We make use of the maximum overlap discrete wavelet transform (MODWT), which has some advantages over the usual discrete wavelet transform: it avoids a downsampling process, unfavourable in some analyses\cite{Percival1994}.} 

 In the input parameters, we choose the least asymmetric wavelet ($LA(4)$) with $8$ scales which provides coefficients that are approximately uncorrelated between scales and reduces the impact of boundary conditions (see the \ref{B-appendix}). For each of the $8$ levels (or wavelet bands), one WV is estimated. We perform an analysis of WV versus scale in a log-log diagram following \citeA{Abry1998} to identify the bandwidth noise at high frequency. In our case, this corresponds to the three first levels of decomposition, which are mostly affected by the data fusion process. The same WV decomposition is applied to the reference time series. The coefficients of the input time series are then rescaled with the WV ratio between the wavelet coefficients of the two time series. No assumption has to be made regarding the noise structure, e.g., it may be a power-law or quantization noise. \textcolor{black}{Intuitively, this is similar to adding the right amount of white noise in the right frequency band and does not distort the underlying signal. Finally, we reconstruct the time series by inserting the rescaled coefficient in the inverse of the decomposition function, i.e. the inverse MODWT. A comprehensive description of the wavelet filter is given
in \ref{B-appendix}.}

\section{Results and Discussions}\label{Section3}

\subsection{Time-Frequency Analysis of the Composite Time Series}

\begin{figure}[htbp!]

\caption{\textcolor{black}{New composite ({\it CPMDF}, orange)  based on merging $41$ years of TSI measurements. For comparison, {\it C3} \protect\cite{Frohlich2006} and {\it C1} \protect\cite{Dudok2017}  are also shown (grey line). A $30$-day running mean of {\it CPMDF} is shown as a yellow/purple dashed line. The orange boxes are associated with the solar minima (SM) for each solar cycle described in Table \ref{Table2}. For context, the monthly sunspot number is also displayed.}
}
\hspace{-6em} 
\includegraphics[width=1.4\textwidth]{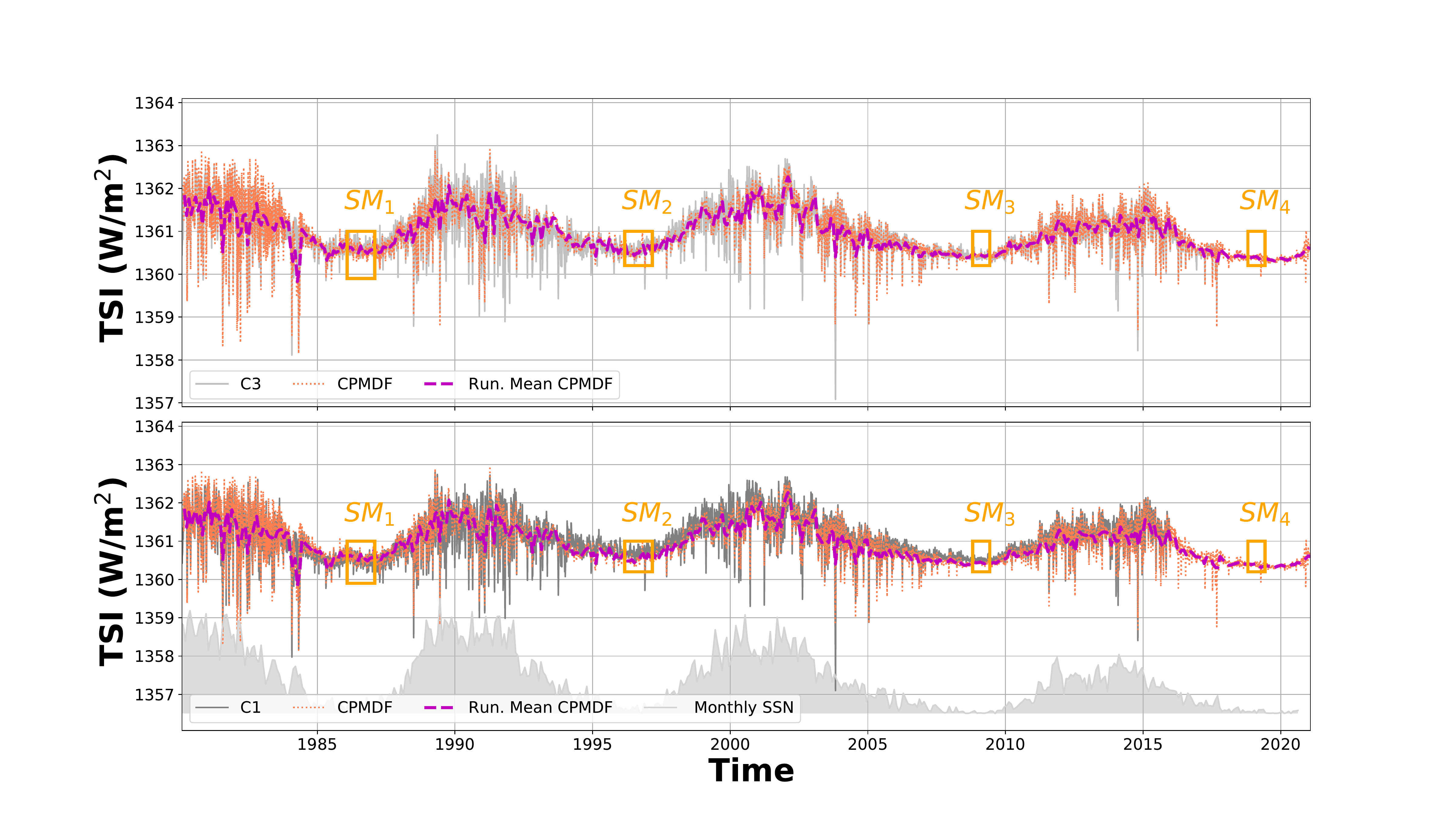}
\label{FigureX2}
\end{figure}

 To perform a time-frequency analysis on the $41$-year TSI composite, we first produce our time series with the $3$-step method. The previous products released by \citeA{Dudok2017}, \citeA{Dewitte2016}  and  \citeA{Froehlich2006} are called respectively {\it Composite 1 (C1)}, {\it Composite 2 (C2)} and {\it Composite 3 (C3)} in the following text. \textcolor{black}{Note that we do not include the ACRIM composite for the sake of clarity in this study. The ACRIM composite \cite{Willson1997} was shown to differ substantially with { \it C1}, { \it C2} and { \it C3}, as discussed in \citeA{Dudok2017}. Thus, it should also differ with our new composite and lead to the same conclusions.} \textcolor{black}{ The new TSI composite is named {\it Composite PMOD- Data Fusion (CPMDF)}. In \ref{B-appendix}, we discuss the composite without applying the wavelet filter. Figure \ref{FigureX2}  displays the composite time series overlaying {\it C1} and {\it C3}}. The estimation of TSI at the solar minimum across the various solar cycles from $1980$ to the present is estimated in Table \ref{Table2}. 
 

\begin{table}[htb!]
\scriptsize
\caption{\textcolor{black}{Estimation of TSI at solar minimum ({\it Minimum}) over the last $41$ years from TSI time series (mean $\mu$ and standard deviation $\sigma$) released by \protect\citeA{Dudok2017} ({\it C1}), by \protect\citeA{Dewitte2016} ({\it C2}) and by \protect\citeA{Frohlich2006} ({\it C3}). The new TSI composite is abbreviated to ({\it CPMDF}). The difference in irradiance between solar minima \textcolor{black}{({\it SM})} from consecutive solar cycles (e.g., $\Delta I_{22/23 -21/22}$) is also displayed with the uncertainties (bold text)} }
\hspace{-5em}
 \begin{tabular}{|c c c  | c Z  c Z  c Z c Z c|}
 \hline
   \multicolumn{3}{|c| }{\multirow{3}{*}{TSI level ($\mu \pm \sigma$ [W/m$^2$])}} & \multicolumn{8}{c}{ Composite Name} & \\ 
    &  &  & \multicolumn{2}{c}{C1} & \multicolumn{2}{c}{C2} & \multicolumn{2}{c}{C3 }  & \multicolumn{2}{c}{CPMDF } & \\
 & &  &  $\mu$ & $\sigma$ & $\mu$ & $\sigma$ & $\mu$ & $\sigma$  & $\mu$ & $\sigma$ & \\
\hline
\multirow{2}{*}{{Solar Cycle 21/22}} & & {\it Minimum ($SM_1$)} & {1360.51} & {0.13} & 1360.39 & 0.13 &  1360.59 &  0.12 & {1360.56}& {0.13} & \\
 &  & $\Delta I_{21/22 -20/21}$ & - &  - &  - & -  & - & -  & - & -& \\
 \hline
 
\multirow{2}{*}{{Solar Cycle 22/23}} & & {\it Minimum ($SM_2$)} & {1360.69} & {0.14} & 1360.46 & 0.16 &  1360.57 &  0.15 & {1360.55}& {0.13} & \\
 & & $\Delta I_{22/23 -21/22}$  & \textbf{0.18} &\textbf{0.25} & \textbf{0.07}  & \textbf{0.28} & \textbf{-0.02} & \textbf{0.27} & \textbf{-0.01} & \textbf{0.26} & \\
 \hline
 
\multirow{2}{*}{{Solar Cycle 23/24}} & &  {\it Minimum ($SM_3$)}& 1360.53 & 0.04 & 1360.44 & 0.04 &  1360.42 &  0.06 & {1360.46} & {0.04} & \\
 & & $\Delta I_{23/24 -22/23}$ & \textbf{-0.16}& \textbf{0.18} & \textbf{-0.02}& \textbf{0.18} & \textbf{-0.15} & \textbf{0.21} &  \textbf{-0.09} & \textbf{0.17} & \\
 \hline
 
\multirow{2}{*}{{Solar Cycle 24/25}} & &  {\it Minimum ($SM_4$)}& - & - & 1360.43 & 0.07 &  - &  - &  {1360.39} & {0.07} & \\
 & & $\Delta I_{24/25 -23/24}$ & - & - & \textbf{-0.01} & \textbf{0.11} & - & - &  \textbf{-0.07} & \textbf{0.11} & \\

\hline
\end{tabular}\label{Table2}
\end{table}

 \textcolor{black}{Note that the solar minima are underlined in Figure \ref{FigureX2} (see {\it SM} - yellow boxes). The solar minimum periods are chosen according to \citeA{Dudok2017} and \citeA{Finsterle2021} by looking at the lowest value in the yearly-averaged sunspot number. We then average the irradiance values over a one-year interval centered on that date to produce the corresponding solar minimum value. The associated uncertainty is the variance of the measurements over the same period.} When comparing the mean difference between the product and our time series over the various solar minima, the new composites agree with {\it C1} at {$0.09  \pm 0.04$} W/$m^2$, {\it C2} at {$0.08  \pm 0.06 $} W/m$^2$,  and {\it C3} at {$0.03  \pm 0.01$} W/$m^2$. The difference is marginal for {\it C1}, {\it C2} and {\it C3}. { Note that {\it C2} is rescaled to the nominal TSI value of $1361$ W/m$^2$ adopted by the IAU 2015, averaged over Solar cycle $23$. By applying this process, the calculated offset is equal to $2.44$ W/ m$^2$, which is due to the calibration based on the absolute level estimated from DIARAD/SOVIM. {\it C1} and {\it C3} are near the nominal TSI value ( below $0.1$ W/ m$^2$ ) averaged over Solar cycle $23$, due to their intrinsic processing.} 
 %
 %
 %
 %

\begin{figure}[htbp!]
\caption{ \textcolor{black}{Power Spectrum Density of TSI { \it C1}  \protect\cite{Dudok2017}, {\it C3}  \protect\cite{Frohlich2006}, together with the new TSI composite produced with the current method {\it CPMDF }}. The $(*)$ means that the time series are shifted by rescaling the amplitude by $-4$  W$^2$ m$^{-4}$ day in the log-log plot. Box $A$, $B$ and $C$ refer to the different sections of the PSD: $A$ is centered on the high frequency ($\sim$ $3$ days) showing the flattening of the PSD; $B$ is the power-law which is mainly due to coloured noise (correlations between $20$ and $6$ days) within the time series; $C$ emphasises the low frequency associated with the stochastic and deterministic parts of the solar cycle and long-term correlations. The dashed lines are the various power-law models when varying the exponent, which are only shown for context. The vertical doted lines (black) mark the frequencies at $11.5$ years, $27$, $9$ and $7$ days (left to right).}
\hspace{-6em}
\includegraphics[width=1.4\textwidth]{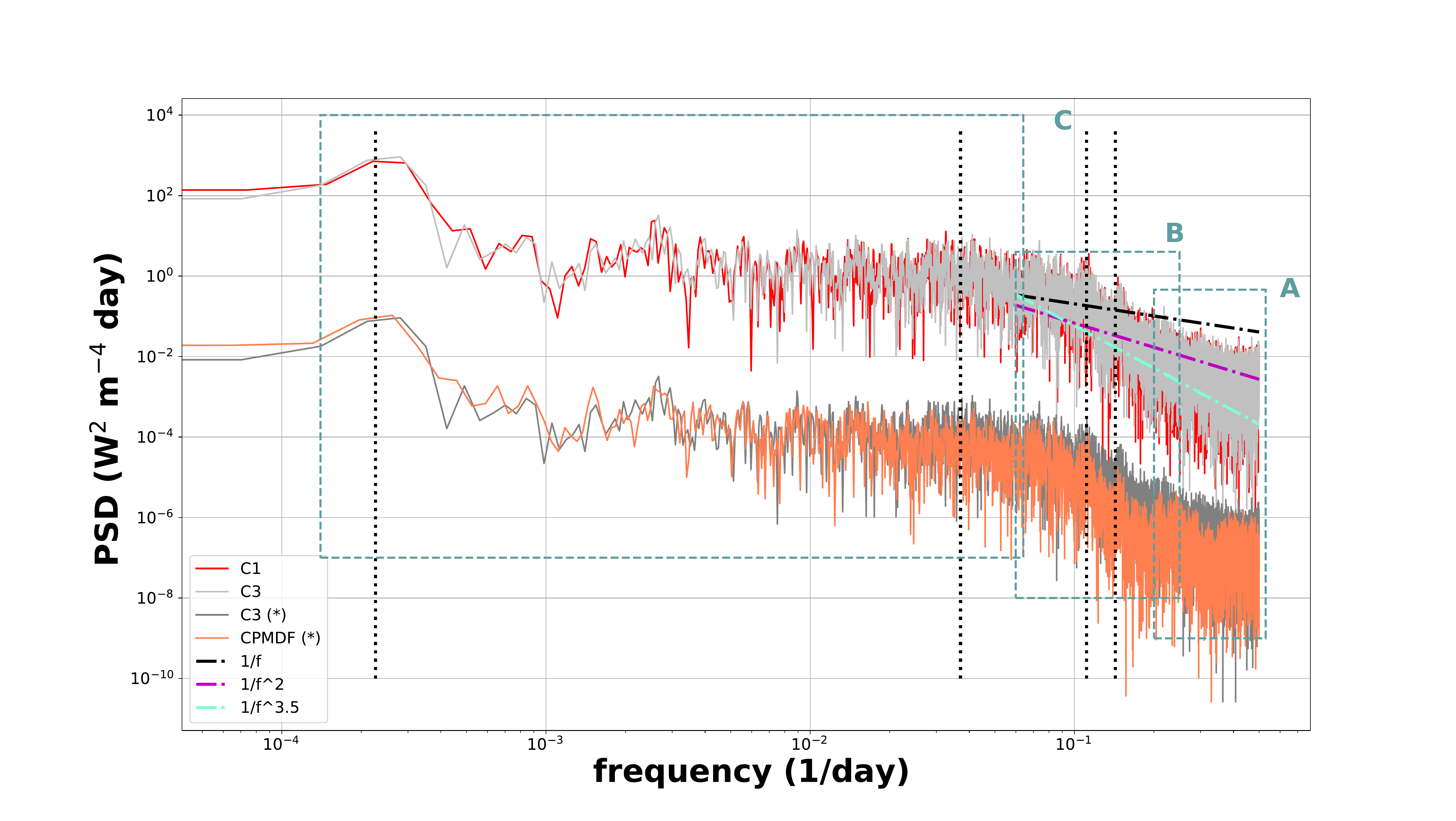}
\label{FigureX3}
\end{figure}

   \textcolor{black}{Now, we analyse the PSD of the composite time series {\it CPMDF} displayed in Figure \ref{FigureX3}}. We can underline the four frequencies ($11.5$ year, $27$, $9$ and $7$ days ) related to solar activity and described by \citeA{Froehlich1997b}. The frequency associated with $11.5$ years is the Schwabe cycle. The quasi $27$-day solar cycle is caused by the sun’s differential rotation (presumably first observed by Galileo Galilei or Christoph Scheiner in the first half of the $17$th century) \cite{Savigny2019}. The spectrum is divided into three areas (i.e. box $A$, $B$ and $C$). 
   The definition of the three areas follows the description of the photospheric activity. The latter associated with granulation, super-granulation and meso-granulation \cite{Andersen1994,Froehlich1997b} generates fluctuations in TSI at different timescales. Due to our daily resolution, frequencies associated with phenomena lasting a few hours or less (i.e. granulation - \cite{Froehlich1997b}) cannot be observed. \textcolor{black}{Note that the instrumental effects due to the recording of the observations cannot be ignored, but it is difficult to decorrelate it from the other noise sources.} 
  
   \vspace{0.5em}
   
    \textcolor{black}{Box $A$ shows a flattening of the curve at high frequencies. The PSD of the products {\it C1} and {\it C3} experience the same flattening at high frequencies. For comparison, \ref{FigureA3} displays the PSD of data recorded by the VIRGO/SOHO radiometer (PMO6V) for the degradation corrected TSI observations on the main channel (VIRGO-A), released in  PMO6v21 \cite{Finsterle2021} at a $1$-minute sampling rate. We can observe that the flattening effect disappears in the sub-daily frequency band.}   \citeA{Shapiro2017} discuss that the high frequencies are associated with the radiometer technical characteristics and satellite movements (i.e. open/close shutter, orbit revolutions). \citeA{Andersen1994} and \citeA{Froehlich1997b} also show that the solar noise flattens in this frequency band. We can then conclude that \textcolor{black}{the flattening curve of the PSD is due} to the low-sampling rate in the TSI composites.

  \vspace{0.5em}
  
\begin{figure}[htbp!]
\caption{ \textcolor{black}{Power Spectrum Density of the difference of composite time series including { \it C3} - {\it CPMDF} , {\it C1} -{\it C3}. The vertical doted lines (black) mark the frequencies at $11.5$ years, $27$, $9$ and $7$ days (left to right).}}
\hspace{-5em}
\includegraphics[width=1.4\textwidth]{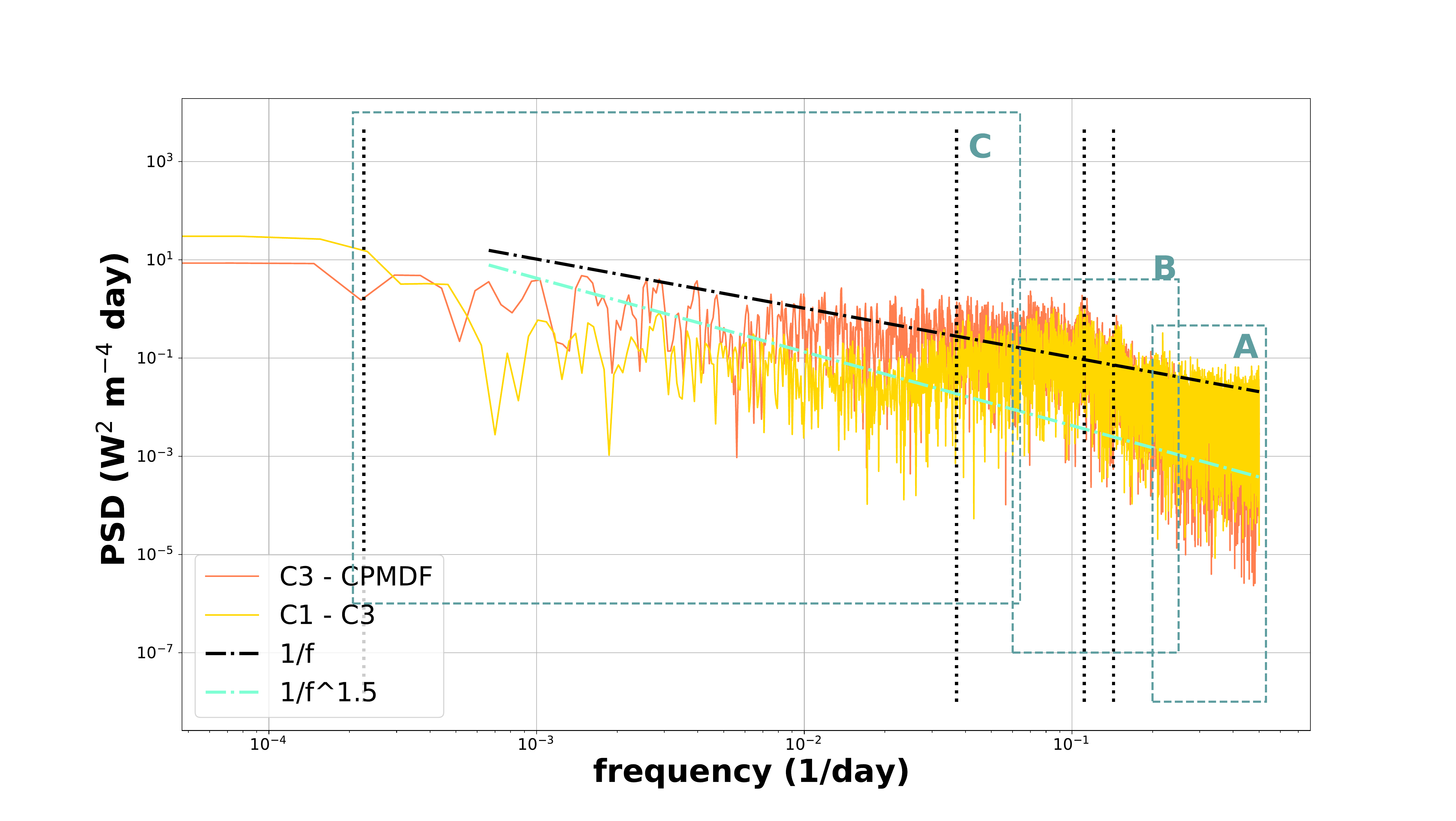}
\label{FigureX4}
\end{figure}

   Box $B$ is the power-law or the frequency ramp (between \textcolor{black}{$0.06$} and $0.25$ $day^{-1}$). This phenomenon is due to the existence of correlations in the observations. It is arguable that this power-law describes the long-term correlations (i.e. over years), due to the ramp spanning frequencies over only a few days ($4$ - $20$ days). Therefore, we can only speculate what underlying process \textcolor{black}{could} generate it. For example, it could be an unknown diffusion process associated with the sun's activity which could be modeled with a specific coloured noise called Mat\'ern process. Nonetheless, the steepness of this ramp shows the degree of correlation or the type of stochastic noise within the time series, by fitting a power-law model such as $S(f) \sim 1/f^\beta$. The exponent $\beta$ defines  the type of coloured noise: flicker noise corresponds to $\beta$  equal to $1$, a random walk to $\beta$ equal to $2$, and white noise with $\beta$ equal to $0$ \cite{Montillet2021}.
   \\ \textcolor{black}{Moreover, in {\it C1}, {\it C2} and {\it C3}, the stochastic noise properties include the correlation from the stochastic part of the solar cycle. In \citeA{Dudok2017}, the authors subtracted various TSI time series from different missions (i.e. {ACRIM1/SMM,  ACRIM2/UARS, ACRIM3/ACRIMSAT, TIM/SORCE}) in order to eliminate the solar cycle, resulting with the stochastic properties only, i.e. a mix of noises between the two instruments. Figure \ref{FigureX4} displays the PSD of the difference of the $41$-year TSI composite. The frequency ramp is mostly attenuated. We can compare its steepness with the various power-law models, hence concluding that the difference composite time series have an exponent $\beta$ within the interval $]1,1.5]$.}  \textcolor{black}{The power-law model is not limited anymore to box $B$ and it includes Box $C$ which advocates for long-term dependencies over years, mainly associated with photospheric activity. This result supports the conclusions in \citeA{Dudok2017}. \textcolor{black}{Note that we cannot exclude the short/long-term correlations due to instrumental effects.}}
 
   \vspace{0.5em}
  
   \textcolor{black}{Finally, box $C$ is associated with the low frequencies ($0.06$ - $0.00015$ $day^{-1}$). They are assumed to be mostly related to the deterministic part of the solar cycle and the long-term correlations (i.e. lasting up to years), including perhaps also some long-term instrumental artifacts.} In the appendices, Figure \ref{FigureA2} shows the spectrum of {\it C1} with and without the solar cycle. \textcolor{black}{To remove the general trend of this cycle,} we subtract the time series using a running mean with a $5$-day window. We clearly see that the low frequencies in box $C$ have the lowest power in the PSD after subtracting the running mean, hence supporting our assumption. In addition, this frequency band also contains some of the coloured noise linked to long-term correlations (over years). Previously, we have discussed the analysis of Figure \ref{FigureX4} when subtracting two TSI composites. We have concluded that the power-law can be extended within Box $C$, highlighting the long-term correlations due to the sun's activity.

 \subsection{Investigating the Solar Minimum Variations}

  Once the $41$-year TSI composite time series is obtained, we can study the existence of variations in the solar minima. There are two approaches: i) the variations between consecutive solar minima, and ii) the \textcolor{black}{global fluctuations (general trend)} over the duration of the time series.
  
    The estimation of the variations between two consecutive solar minima is challenging based on the PSD analysis of the composite time series. In order to be statistically robust, one needs to take into account the long-term correlations generated by the coloured noise. 
    Therefore, we follow the same methodology as in \cite{Dudok2017}, where we differentiate the estimated irradiance at solar minima between four consecutive cycles. The results are shown in Table \ref{Table2} (see e.g., $\Delta I_{22/23 - 21/22}$). Overall, the fluctuations of the solar minima between Solar Cycle  $21/22$ and $22/23$ ($\Delta I_{22/23 - 21/22}$) do not agree between the composites. For example, the difference is positive  for the {\it C1} and {\it C2}, whereas it is negative for the other composites. This disagreement could be due to the processing of the TSI observations for the first missions (e.g., {HF/NIMBUS-7 ERB, ERBE/ERBS}) discussed in Section \ref{Section2.1}. \textcolor{black}{The fluctuation between the other solar cycles (i.e  Solar Cycle $22/23$,  $23/24$,  $24/25$) is more homogeneous in terms of the sign value (i.e. negative for all of them). The averaged value  is $-0.10  \pm 0.37$ W/m$^2$ and $-0.04  \pm 0.15$ W/m$^2$ for  $\Delta I_{23/24 - 22/23}$ and $\Delta I_{24/25 - 23/24}$, respectively.} 
    \textcolor{black}{However, any trend or large-scale fluctuation is downplayed by the large uncertainties associated with the difference between solar minima - up to $4$ times the value.} 
    
    For the study of the global \textcolor{black}{trend (large-scale fluctuation) over the entire TSI composite time series}, our approach is inspired by the estimation of a tectonic rate in geodetic time series \cite{Davis2012,Montillet2020}. The problem is formulated  into a joint estimation of functional and stochastic models.
  The functional model is composed of two terms a linear trend and a periodic signal with $4$ frequencies ($11.5$ years, $27$, $9$ and $7$ days) based on our PSD analysis. Because of the flattening experienced by the TSI composite time series at high-frequency (see above discussions - Box $A$), the use of the  General Gauss-Markov model (GGM) with white noise is appropriate in order to model the frequency ramp feature of the PSD (i.e. Box $B$). The justification of the model and the estimation of the parameters (using  a MLE) are described in the appendices. 
  Now, all the composite time series experience a much lower solar minima at the end of cycle $21$ around $1986$, it renders the fitting of the deterministic part of the solar cycle difficult with a periodic signal. That is why we perform this study by splitting the composite time series \textcolor{black}{into two time intervals} $1980$-$2021$ (including cycle $21$) and $1987$-$2021$ (starting at cycle $22$). The analysis of the functional model fitting the residual shows that the model fits best when using the period $1987$- $2021$. This confirms our previous study where the exclusion of solar cycle $21$ allows the trend between the difference in solar minima to be debated. 
 \begin{table}[htb!]
 \scriptsize
\caption{\textcolor{black}{Estimation of the linear trend (mean $\mu$ and uncertainty $\sigma$) via MLE using the GGM model together with white noise for the TSI composite time series released by  \protect\citeA{Dudok2017} ({\it C1}), \protect\citeA{Dewitte2016} ({\it C2}),  \protect\citeA{Frohlich2006} ({\it C3}) and by applying (i.e. {\it CPMDF}) the wavelet filter. Two different time periods are chosen. $yr$ means year}} 
\hspace{0em}
 \begin{tabular}{||c|cZ|cZ||} 
 \hline
   TSI level ($\mu \pm \sigma$ [W/(m$^2$yr)]) &   \multicolumn{2}{c|}{ } &  \multicolumn{2}{c||}{ } \\
  Period  &     \multicolumn{2}{c|}{1980 - 2021}       &  \multicolumn{2}{c||}{1987 - 2021 }  \\
  Amplitude Solar Trend & $\mu$ & $\sigma$  & $\mu$  & $\sigma$  \\ 
 \hline
 C1  & {-0.004} & {0.006} & {-0.001} & {0.009} \\
 C2 & -0.001 & 0.006 & -0.003 & 0.007  \\ 
 C3 & -0.009 & 0.006  & -0.011 & 0.009   \\
 CPMDF & {-0.014} & {0.004}   & {-0.010} & {0.006} \\
  \hline
\end{tabular}\label{Table3}
\end{table}

  Table \ref{Table3} displays the results for each TSI composite time series. {\it C3} has the largest trend for  both periods $1980$ - $2021$ and $1987$ - $2021$ compared with {\it C1} and {\it C2}.  \textcolor{black}{The trend of the new product {\it CPMDF} is larger than the previous product for the period $1980$ - $2021$. However, it is the same order of magnitude as {\it C3} for the second period.}
  
  Averaging the estimated trend for all the previous products gives $-0.005  \pm 0.003$ W/(m$^2 yr$) and $-0.005  \pm 0.004$ W/(m$^2 yr$) for the periods $1980$ - $2021$ and $1987$ - $2021$, respectively. \textcolor{black}{When we include the new product, the average trend is $-0.004  \pm 0.003$ W/(m$^2 yr$) and $-0.004  \pm 0.004$ W/(m$^2 yr$) for the same periods. The overall estimate, using the results from all the TSI composite time series and both periods, is equal to $ \sim -0.004$ $\pm$ $0.004$ W/(m$^2 yr$).}
  %
  %
  Overall for each product, the uncertainty associated with the estimated trend is large, mostly larger than the amplitude of the trend. This result means that the estimated amplitude is statistically insignificant: the stochastic properties of the composite time series are likely the source of the variations. This corroborates the previous results based on the estimation of the variations between two consecutive solar minima. Both are showing the same pattern after cycle $21$. Note that these conflicting decadal trends exhibited by the previous TSI composites ({\it C2} and {\it C3}) are discussed by \citeA{Yeo} using proxy data. 
  Furthermore, most of the estimated amplitudes are negative. This has a certain significance related to the solar noise which describes the solar activity, hence meaning that over the last $41$ years there has been a slowly decreasing solar activity.  This result is supported by several studies focusing on the forecast of the sun's activity over the next $80$ years \cite{Beer2013,VELASCOHERRERA2015221}.
  
\section{Conclusions}
     \label{Section4} 

{ We have merged  $41$ years of satellite observations using data fusion in order to produce a TSI composite time series, which can be used to study the solar cycle modulation and the Earth's energy budget. We have performed a time-frequency comparison of our new TSI composite with previous releases, including the TSI community consensus time series {\it C1}.} The results show  {that the mean value of the difference over the solar minima is below the $1$ sigma confidence interval of $0.2$ W/m$^2$, i.e. a maximum of $0.09$ W/m$^2$ for {\it C1} and a minimum of $0.03  \pm 0.01$ W/m$^2$ for {\it C3}}. \textcolor{black}{In terms of comparing the frequency spectrum, we observe a flattening at high frequencies for all products which is linked to the various instrumental noises and the low sampling rate ($1$ day). We expect that future TSI composite time series including observations recorded from future missions will be produced with a higher time resolution ( i.e. hourly, sub-hourly) in order to include the meso-granulation, granulation and p-modes frequency bands.}

  \textcolor{black}{Secondly, the power spectrum experiences a power-law between $4$ and $20$ days which could correspond to an unknown diffusion process. When removing the solar cycle by differencing two TSI composite time series, a frequency ramp (power-law noise) is observed over the whole frequency band. The power-law exponent varies within $[1,2]$. It highlights the presence of long-term correlations from solar noise and perhaps instrumental noise.}
%

    Finally, our approach permits the estimation of a trend in the $41$-year composite TSI time series which could reflect variations in the solar activity. The analysis of the irradiance difference ($\Delta I$) estimated at two consecutive solar minima  in order to detect a trend is inconclusive due to large uncertainties. \textcolor{black}{Our results using a joint inversion of both a functional and stochastic noise models, show that the estimated amplitude is below  $ \sim -0.004$ $\pm$ $0.004$  W/(m$^2 yr$) based on the analysis of all the $41$-year TSI composite time series used in this study.} This number is not statistically robust due to the large uncertainties. Therefore, it is impossible to reach any conclusions about the existence of a linear trend in the TSI composite time series. Any visual effects or short-term trends are most likely related to the coloured noise rather than a physical phenomenon generated by the sun's activity, corroborating previous discussions \cite{Dudok2020} and supporting recent analysis \cite{Werner2021}. 

% ... 
\acknowledgments
\textcolor{black}{We acknowledge the life-long dedication to the TSI community of the late Dr. Claus Fröhlich in memoriam as a former director of PMOD/WRC, PI of SOHO/VIRGO and his invaluable contribution to the analysis of TSI observations. Among them, he produced the first reconstruction of the TSI composite in $2006$ which is also used in this work (i.e. {\it C3}). 
Dr. J.-P. Montillet, Dr. W. Finsterle, Dr. M. Haberreiter and Prof. W. Schmutz gratefully acknowledge the support from the Karbacher-Funds. Dr. G. Kermarrec would like to acknowledge the Deutsche Forschungsgemeinschaft under the project KE2453/2-1 which permitted the development of the wavelet filter to analyze correlated noise, opening the door for further studies related to laser observations. The authors thank the anonymous reviewers whose comments/suggestions helped improve and clarify this manuscript.} 

\section*{\textcolor{black}{Open Research}}

\noindent \textcolor{black}{ The new composite {\it CPMDF} can be obtained from the open archive repository $www.astromat.org$ \cite{MontilletData2022}. It is also presented on $www.pmodwrc.ch/en/?s=TSI+Composite$ (last accessed 07 June 2022) for additional information. The TSI composite {\it C1} is available for downloading at $http://www.issibern.ch$ $/teams/$ $solarirradiance$ (last accessed 07 June 2022). The data related to the monthly/daily mean sunspot numbers are retrieved from $http://www.sidc.be/$ $silso/datafiles$ (last accessed 07 June 2022). {{\it TIM /SORCE},  {TCTE/TIM} and {\it TIM/TSIS} time series} are downloaded from   $https://lasp.colorado.edu$ $/home/sorce/data/tsi-data$ (last accessed 07 June 2022). {\it PREMOS (v1)}  can be accessed at $http://idoc-picard.ias.u-psud.fr/sitools/client-user/Picard/project-index.html$ (last accessed 07 June 2022). {\it PMODv21a} is available at $https://www.pmodwrc.ch/en/research-development/space/soho/$ (last accessed 07 June 2022).}

%
%
%
%
\clearpage
\appendix
\renewcommand\thefigure{\thesection.\arabic{figure}} 

\section{Model Descriptions for Estimation of the Linear Trend} \label{A-appendix}

Following the discussion in Section \ref{Section3},  the stochastic noise model of the TSI time series is described by the variance \cite{Williams2004}:
  
\begin{equation}\label{Secondeqbis}
E\{\mathbf{\psi}^{T}\mathbf{\psi}\} = \sigma_{wn}^2\mathbf{I} + \sigma_{pl}^2 \mathbf{J}(\beta)
\end{equation}
  
\textcolor{black}{where $E\{.\}$ is the expectation operator.} The vector $ \mathbf{\psi} = [\psi(t_1), \psi(t_2), ..., \psi(t_L)]$ is \textcolor{black}{a multivariate continuous-time stochastic process}. At each \textcolor{black}{time step}, we define $\psi(t_i) = \psi_{wn}(t_i) + \psi_{pl}(t_i)$, with $\psi_{wn}(t_i)$ and $\psi_{pl}(t_i)$ the white Gaussian noise (zero mean) and the coloured noise (or power-law noise) sample respectively.  $T$ is the \textcolor{black}{transposition} operator,  $\mathbf{I}$ the identity matrix, $\sigma_{pl}^2$ the variance of the power-law noise and $\mathbf{J}(\beta)$ the covariance matrix of the power-law noise ($\beta$ $>$ $0$). The definition of $\mathbf{J}$ depends on the assumptions of the type of coloured noise (e.g., flicker, random-walk).

 The functional model  $s_0(t)$ (at epoch $t$) is based on the polynomial trigonometric method   \cite{Williams2004,Montillet2020}.
\begin{equation}\label{firsteq}
s_0(t) = at+b+\sum_{j=1}^N (G_j\cos(D_jt) + E_j\sin(D_jt))
\end{equation}

with $a$ and $b$ the coefficients of the linear rate; the deterministic part of the solar cycle is modeled by a sum of $\cos$ and $\sin$ functions with coefficients $G_j$ and $E_j$. Note that $D_j$ ($2\pi fq_j$) and $fq_j$ are different frequencies (e.g., $11.5$ years, $27$, $9$ and $7$ days) which are determined by analyzing the frequency spectrum of the TSI composite time series (see Section  \ref{Section3}). We perform a joint estimation of the functional and stochastic models  based on a MLE. To recall \citeA{Bos2020}, the log-likelihood for a time series of length $n$ can be rewritten as:
\begin{equation}\label{Lofun}
	\ln(Lo)= -\frac{1}{2}\left[n\ln(2\pi) + \ln(\det({\bf C})) + ({\bf x_0} - {\bf A}{\bf z})^T{\bf C}^{-1}({\bf x_0} - {\bf A}{\bf z})\right]
\end{equation}

This function must be maximised. Assuming that the covariance matrix ${\bf C}$ is known, then it is a constant and does not influence finding the maximum. \textcolor{black}{${\bf C}$ is equal to $E\{\mathbf{\psi}^{T}\mathbf{\psi}\}$} as defined by Eq. \eqref{Secondeqbis}. The term \(({\bf x_0} - {\bf A}{\bf z})\) \textcolor{black}{represents} the TSI observations minus the fitted model. Note that \(( {\bf A}{\bf z})\) is the matrix notation of $s_0$. The last term can be written as \({\bf x}^T{\bf C}^{-1}{\bf x}\) and it is a quadratic function, weighted by the inverse of matrix \(\bf C\). To select the functional model of the solar signal, and therefore estimate the associated parameters, we have formulated the assumptions in Section \ref{Section2} and the time-frequency analysis in Section \ref{Section3}. The value of $n$  is here equal to the number of observations in the TSI composite time series ($\sim 15330$ observations).
%
%


    The particularity of the TSI composite frequency spectrum has been discussed above, and in particular its flattening at high frequencies. Therefore, a simple power-law noise model as described in Section \ref{Section2} with the covariance $\mathbf{J}(\beta)$ is not appropriate for our ML estimation. Instead, we use the Generalized Gauss-Markov (GGM) noise model which has the advantage of flattening at high frequencies. The PSD of the GGM noise is defined by \citeA{Bos2020} as:

\begin{equation}\label{GGMmodel}
    S(f) = \frac{2 \sigma^2}{f_s^2} \Bigg[ 1+ \phi^2 -2\phi \cos{(2 \pi \frac{f}{f_s})}\Bigg ]^{-\beta/2}
\end{equation}

 where $\phi$ is an important parameter to decide when the flattening occurs in the PSD. In our study with TSI time series, \textcolor{black}{we have fixed $\phi$ to $1.0699$ following the recommendation of \citeA{He2019}}. Also from Eq. \eqref{GGMmodel}, if $\phi$ is set equal to $1$ then the PSD is similar to an approximation of the power-law model. For more information and discussions about this model, we invite the reader to refer to \citeA{Bos2014}. Note that we use the Hector package to do the joint model estimation.


\clearpage
\section{Description of the Wavelet Filter} \label{B-appendix}
This section comprehensively describes the wavelet filter proposed in Section \ref{Section2.3}. We discuss the steps to correct some of the long-range dependency introduced by the data fusion process to generate the final TSI {\it CPMDF}. In this section, let us call {\it CPMDFa}, the composite time series before applying the wavelet filter. 

\subsection{Methodology}

\textcolor{black}{The method for rescaling the high-frequency power of CPMDF can be divided into 4 steps:}

\begin{enumerate}
    \item The MODWT decomposition : discrete versus maximum overlap discrete wavelet transform
  
\textcolor{black}{ The first step of the wavelet filter is the wavelet decomposition of {\it CPMDFa} itself. To that end, the time series is usually broken down into a scaled and shifted version of a chosen mother wavelet. Because the sample size of the orthogonal discrete wavelet transform (DWT) is limited to a power of $2$, the number of scaling and wavelet coefficients at each level of resolution decreases by the same factor.} Unfortunately, this results (i) in a loss of information as the level of decomposition increases, as well as (ii) in the introduction of ambiguities in the time domain. These effects are unwanted when performing a scale-by-scale variance analysis. We circumvent the drawback of the DWT by using the 
maximal overlap discrete wavelet transform (MODWT, \cite{Cornish2006}), which  carries out the same steps as the DWT without a sub-sampling process. \textcolor{black}{Mathematically, the MODWT is a convolution operation that can be formulated as circular filter operations of the original time series using $2$ quadrature mirror filters, Therefore, the decomposition at each scale can be understood as a bandwidth filtering of the original time series in the frequency domain with various high and low-pass filters.}
The MODWT is known as a shift-invariant wavelet transform. It is a highly redundant version of the DWT and is considered ideal for time-series analysis, as it accommodates any sample size. In this contribution, we use the MATLAB wavelet package from Mathworks ($https://ch.mathworks.com/$).
 
 
     Whereas low scales of the MODWT decomposition are related to long periodic behaviour, high scales focus on brief phenomena. \textcolor{black}{The proposed wavelet filter aims to rescale the power of the high frequencies of {\it CPMDFa}, which are filtered out by the data fusion, and recompose the obtained time series: the choice of the wavelets should take this application into consideration.} Here, we follow the work of \citeA{Cornish2006} who proposed to use the least asymmetric (LA) wavelets. These wavelets exhibit near symmetry about the filter midpoint which allows a good alignment of the reconstruction with the original time series by circularly sifting the coefficients. \textcolor{black}{More specifically, we use the LA(4): this wavelet has a nearly linear phase response and is optimal for reconstruction. Compared to the Haar wavelet, it has less leakage, which makes it more appropriate to rescale the variance of the chosen levels of decomposition (see step 2). Here the band-pass filtering is more accurate and allows a better control over the rescaled levels of decomposition at high frequencies. We further note that the 4 vanishing moments produce wavelet coefficients vectors that are nearly stationary; this is favorable to analyse the variance. We then chose to decompose the signal into $8$ levels, which we justify by our specific focus on the high frequency domain. The low frequencies do not need particular attention and should be kept intact with the purpose of not losing information from the data fusion.}
 
 \item The wavelet variance (WV) decomposition
 
In a second step, we compute the variance of each decomposition level time series, on a scale-by-scale basis. The variances are called the wavelet variances (WV). \textcolor{black}{The WV can be interpreted as the variance of a process after filtering by a wavelet bandpass filters \cite{Percival1994}. Our method to rescale the WV is based on an unbiased wavelet estimator developed by \citeA{McCoy1996} and \citeA{Abry1998} for correlation analysis.}
The WV multiscale analysis provides an efficient and highly robust estimator of the fractal parameter of a process. We recall that fractal processes have a power of the form $1/f^{\beta}$  for a range of frequency $f$  close to $0$.
 \textcolor{black}{It can be shown that the WV versus its scales have a logarithmic linear relationship for such processes: the slope in a $log_{2}$-diagram is related to the power law of the process and can be estimated by ordinary least-squares \cite{Abry1998}. Here, we propose to use the WV to eliminate the correlations at high frequencies induced by the windowing used to fuse TSI from different datasets.}
 
 \item Rescaling the WV

  The procedure is inspired by the work of \citeA{Guerrier2013} on composite stochastic processes and is based on the standardized distance between the WV of a reference process and that under consideration. When dealing with a mix of unknown bandwidth fractal noises, the scales at which a given noise is present have to be identified in a first step in the $log_{2}$-diagram WV versus scales. \textcolor{black}{As an example, a white noise exhibits a slope of $-1$, flicker noise has a slope of $0$ and, a random-walk has a slope of $1$. An alignment of the WV versus a given number of scales is linked with a bandwidth fractal noise. Here, we want to identify the correlated noise introduced by data fusion, which was shown to be found in the high frequency domain. We illustrate our decorrelation or rescaling procedure using { \it C3}. Figure \ref{figLast8}.A is a $log_{2}$-diagram showing the WV versus scales. Our reference time series is marked with blue dots, together with the fused time series before (green) and after (pink) filtering. The WV of both time series lies on a straight line. For the reference time series, the slope approximately corresponds to a coefficient equal to $-3$ in the WV $log_{2}$-diagram. The first WV contains most probably an additional WV component as it is slightly over the line drawn from scale $2$ and $3$. From scale $4$, the WV spectrum clearly changes its shape, which corresponds to the Mat\' ern process (saturation at low frequency). This visual analysis of the $log_{2}$-diagram shows that the sought after high frequency correlated noise is present between the scales $1$ and $3$.}
  \textcolor{black}{Here we propose to rescale the WV of {\it CPMDFa} in order to fit the reference one (i.e. { \it C3}). The resulting WV spectrum is shown in the green dots in Figure \ref{figLast8}.A. Intuitively, we have re-introduced high frequency noise to obtain the reference decay of the WV spectrum (or similarly that of the power spectral density).}

 \item  \textcolor{black}{Recomposition of the rescaled TSI}

 \textcolor{black}{ Finally, we recompose the filtered time series by using the inverse function inserting our new WV values. Note that we use the function called Inverse MODWT, abbreviated as IMODWT, which is also included in the MATLAB wavelet package. Our new time series is comparable to both { \it C1}, { \it C2} (after shifting of $\sim -2.44$ W/m$^2$), { \it C3} and  {\it CPMDFa}, well within the interval of confidence of the reconstruction ($0.2$ W/m$^2$) as discussed in Section \ref{Section3}. Note that the MODWT and IMODWT require $O( n.log_2 .n)$ multiplications \cite{Percival2000}.}

  \end{enumerate}  
  
   \subsection{\textcolor{black}{Discussion of the Effect of Applying the Wavelet Filter}}
   
\begin{figure}
\includegraphics[width=0.9\textwidth]{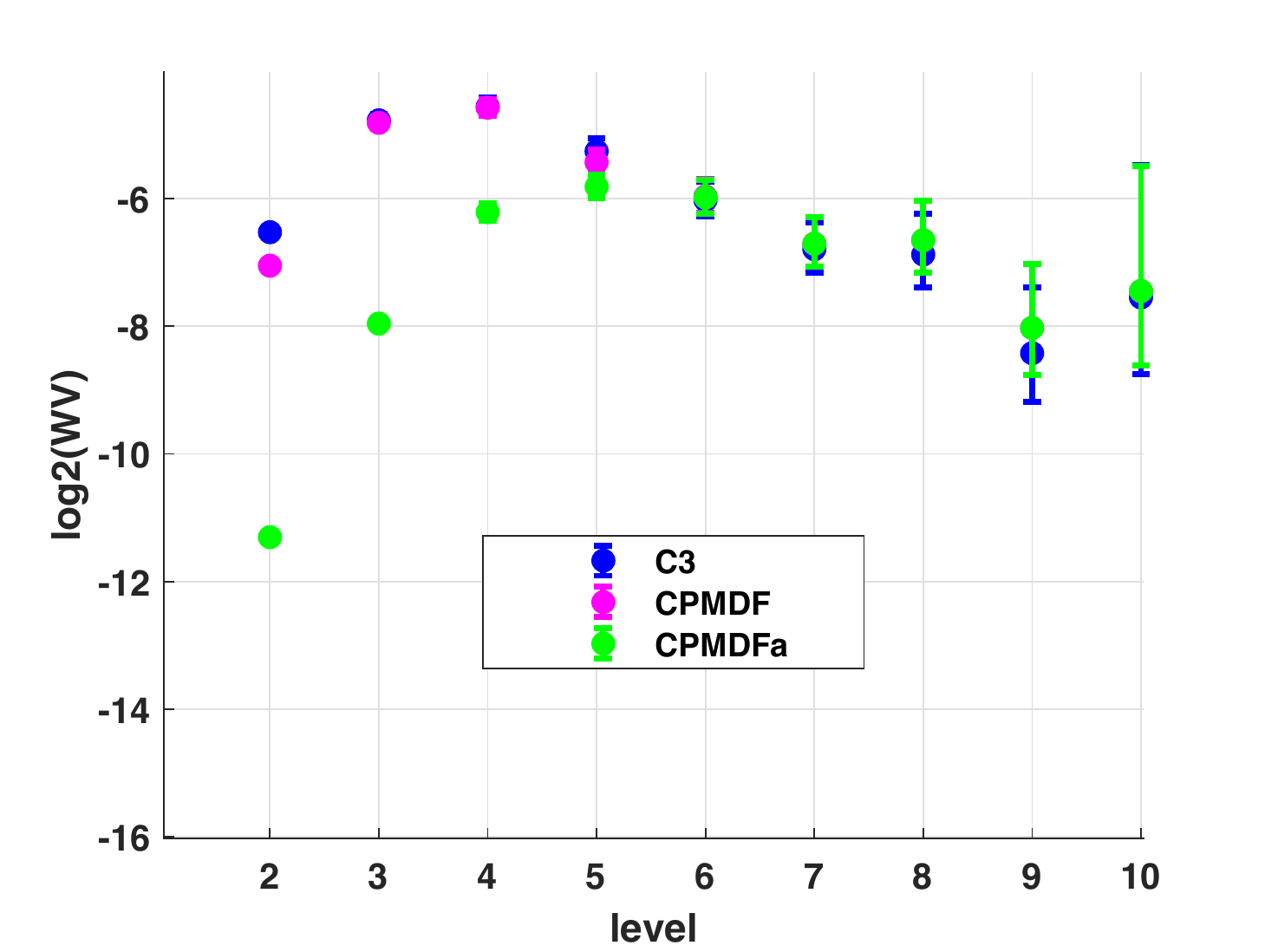}\\
\includegraphics[width=0.9\textwidth]{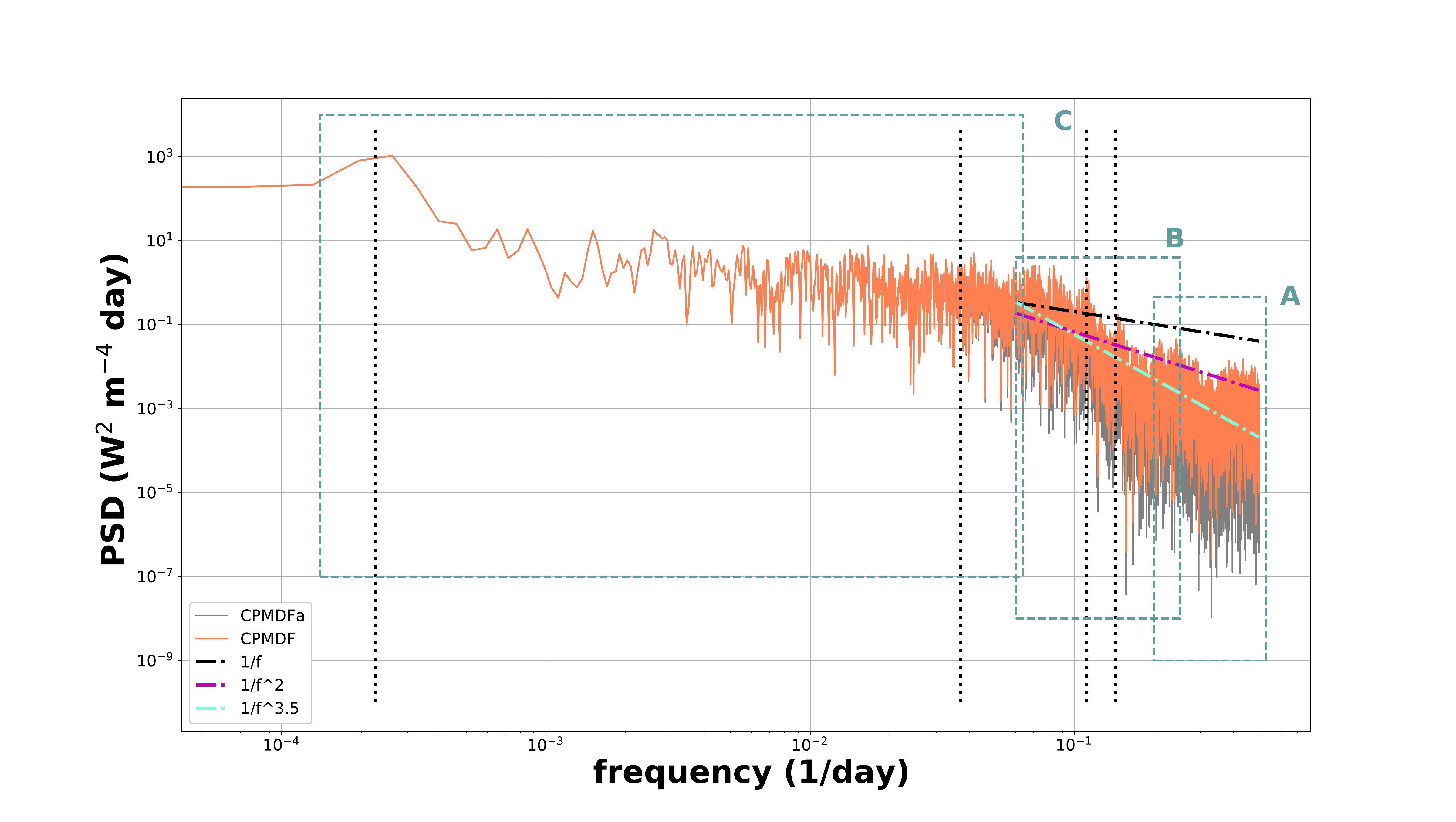}
\caption{  (Top) \textcolor{black}{Wavelet Variance decomposition of { \it C3} (blue), {\it CPMDFa} (green) and the filtered time series (pink). The X-axis is the level of decomposition, whereas the Y-axis is the WV value on a log scale. (Bottom) Effect of applying the wavelet filter in the frequency domain: without {\it CPMDFa}  (grey) and with {\it CPMDF} (orange).}}
\label{figLast8}
\end{figure}

   \textcolor{black}{It is worth showing the impact of applying the wavelet filter on the TSI time series composite. Figure \ref{figLast8}.B displays the PSD of the time series before and after the filter.  The difference can only be seen in Box  $B$. The steepness of this ramp is more accentuated for {\it CPMDFa} than in {\it CPMDF}. The power-law is between $2.5$ and $3$ for the previous releases {\it C1}, {\it C2}, {\it C3} and {\it CPMDF}, but between $3$ and $3.5$ for {\it CPMDFa}. This increase steepness is a weakness of the data fusion process as discussed in Section \ref{NewSection2}, which can smooth the short-term and long-term correlations. It is a nonlinear effect of the input parameters (e.g., inducing points). Increasing this number decreases the steepness to a certain extend. This problem is intractable when using GPs with a very large number of inducing points (e.g., $2000$ points) due to computational complexity. Nevertheless, the implementation of a wavelet filter has shown that we can efficiently reconstruct the high frequency bandwidth, hence having a PSD comparable with previous products and without the cost of increasing the processing time.} 
  


\clearpage
\section{Additional Figures and Remarks} \label{C-appendix}


\subsection{\textcolor{black}{Remarks on the Inducing Points}}

\begin{figure}[htbp!]
\caption{Time series of {\it C3} \protect\cite{Frohlich2006} and the sub-time series in Box $8$ (see Figure \ref{FigureX1}) fusing VIRGO/SOHO, {ACRIM3/ACRIMSAT} and TIM/SORCE using various numbers of inducing points ($500$, $2000$). The sub-time series are aligned on {\it C3}.}
\hspace{-5em}
\includegraphics[width=1.4\textwidth]{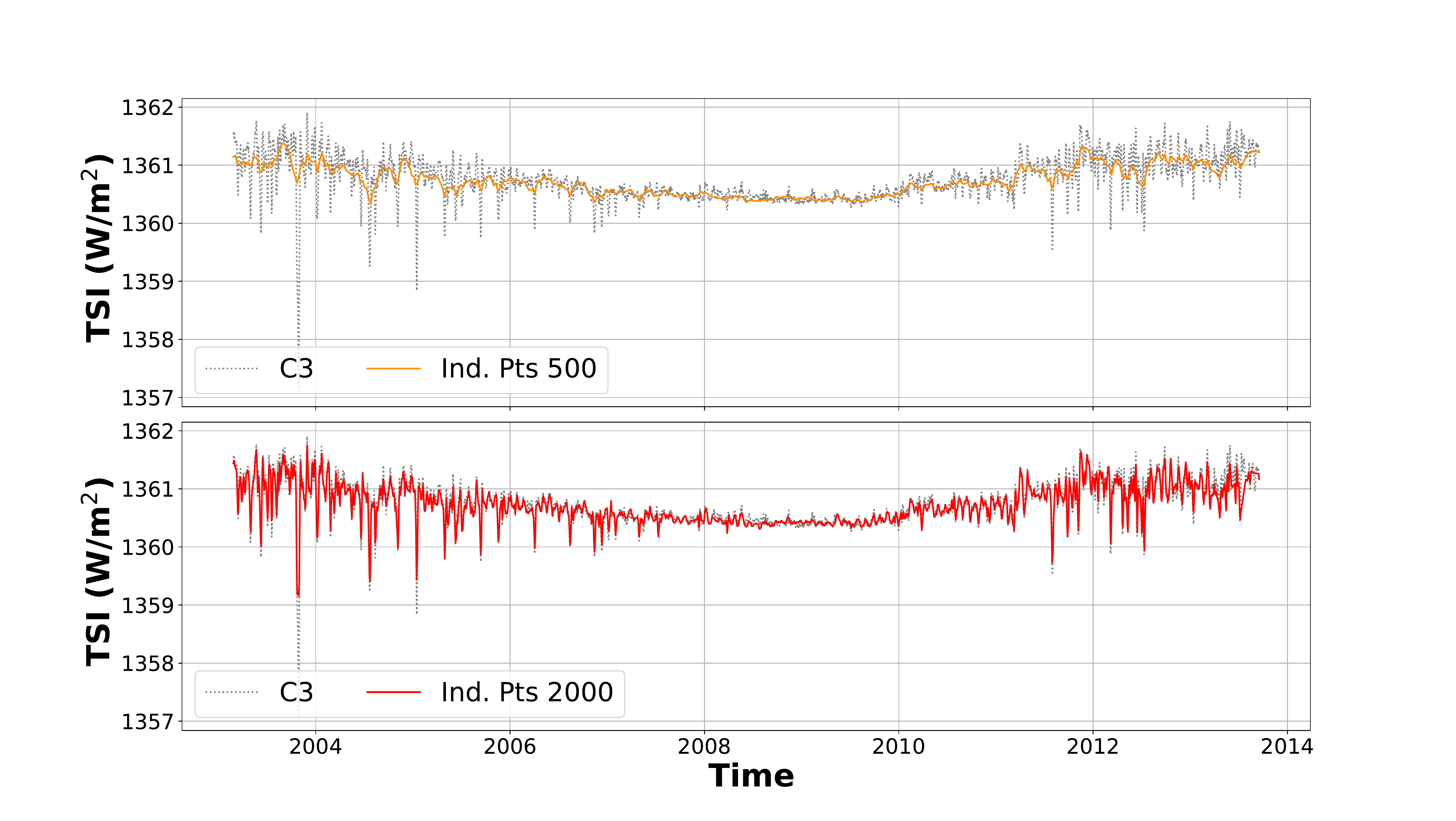}
\label{FigureA0a}
\end{figure}

\begin{figure}[htbp!]
\caption{Power Spectrum Density of TSI {\it C3} \protect\cite{Frohlich2006} and the sub-time series in Box $8$ fusing VIRGO/SOHO, ACRIM3/ACRIMSAT and TIM/SORCE using various numbers of inducing points ($500$, $1800$).
 Box $A$, $B$ and $C$ refer to the different sections of the PSD: $A$ is centered on the high frequency ($\sim$ $3$ days), which shows flattening of the PSD; $B$ is the power-law which is mainly due to coloured noise (correlations between $20$ and $6$ days) within the time series; $C$ emphasises the low frequency associated with the stochastic and deterministic parts of the solar cycle and \textcolor{black}{long-term correlations, as well as instrumental artifacts.} The dashed line is the power-law model when varying the exponent (shown for context). }
\hspace{-5em}
\includegraphics[width=1.4\textwidth]{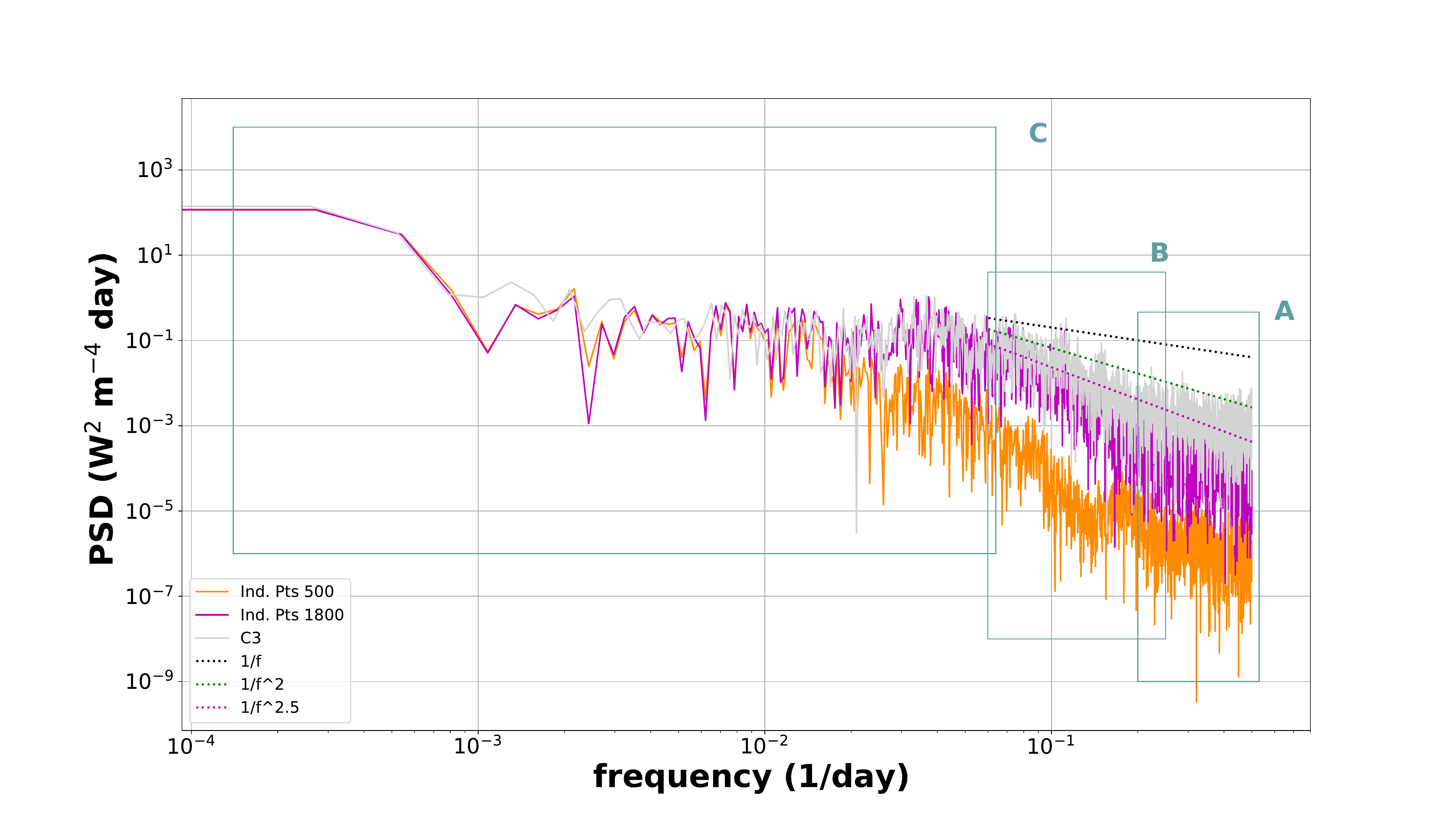}
\label{FigureA0b}
\end{figure}

\textcolor{black}{Figure \ref{FigureA0a} shows the variations in both time and frequency  when fusing VIRGO/SOHO, TIM/SORCE and ACRIM3/ACRIMSAT (box $8$ in Figure \ref{FigureX1}). Figure \ref{FigureA0b} displays the associated power spectrum density (PSD). It is difficult to find an optimal number, because above $1500$ points the time fluctuations do not show many differences (i.e. continuity of the spectrum, amplitude of the frequencies associated with the solar cycle).}

\subsection{\textcolor{black}{Additional Figures}}
\begin{figure}[htbp!]
\caption{Power Spectrum Density of TSI {\it C3} \protect\cite{Frohlich2006} with and without partially removing the solar cycle via a running mean (with a $5$-day window). Boxes $A$, $B$ and $C$ refer to the different sections of the PSD. The dash-dotted lines are the various power-law models when varying the exponent. The vertical dotted lines (black) mark the frequencies at $11.5$ years, $27$, $9$ and $7$ days. The purple dash-dotted lines highlight the change of power in box $C$ before and after removing the solar cycle. }
\hspace{-5em}
\includegraphics[width=1.4\textwidth]{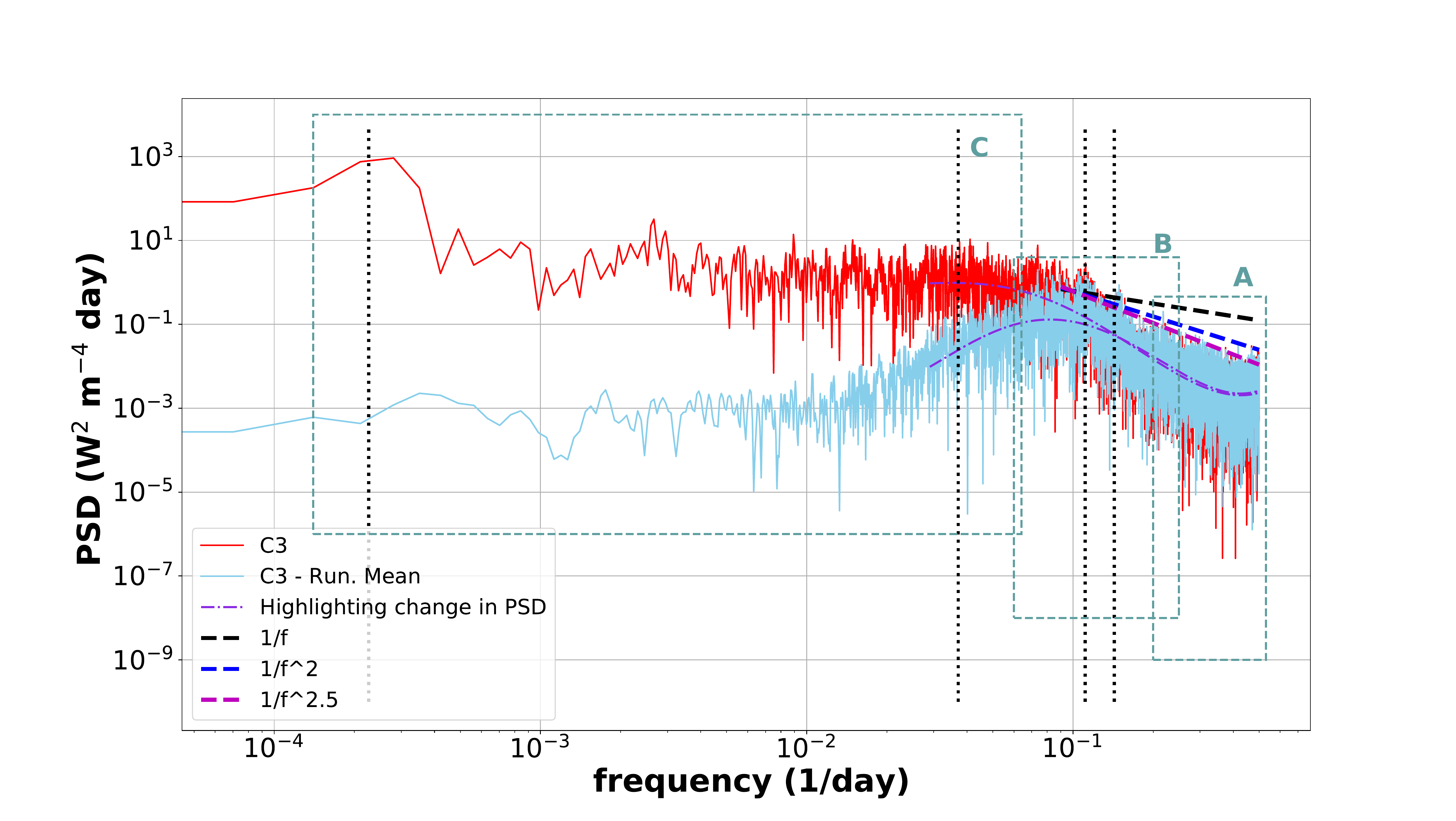}
\label{FigureA2}
\end{figure}
\begin{figure}[htbp!]
\caption{Power Spectrum Density of various products: VIRGO/SOHO with degradation correction (VIRGO-A) from PMO6v21 with $1$-minute and daily sampling rate.  {\it C3} is the TSI composite produced by \protect\citeA{Frohlich2006}. {\it CPMDF} is our new TSI composite including the wavelet filter. Note that for clarity the different spectra have been rescaled by multipying them by $-6$, $-6$, $-4$ and $-8.5$ W$^2$ m$^{-4}$ day following the order of the legends from the top.  Boxes $A$, $B$ and $C$ refer to the different sections of the PSD. The dash lines are the various power-law models when varying the exponent (shown for context). The vertical dashed line emphasizes the $11.5$ year peak.
}
\hspace{-5em}
\includegraphics[width=1.4\textwidth]{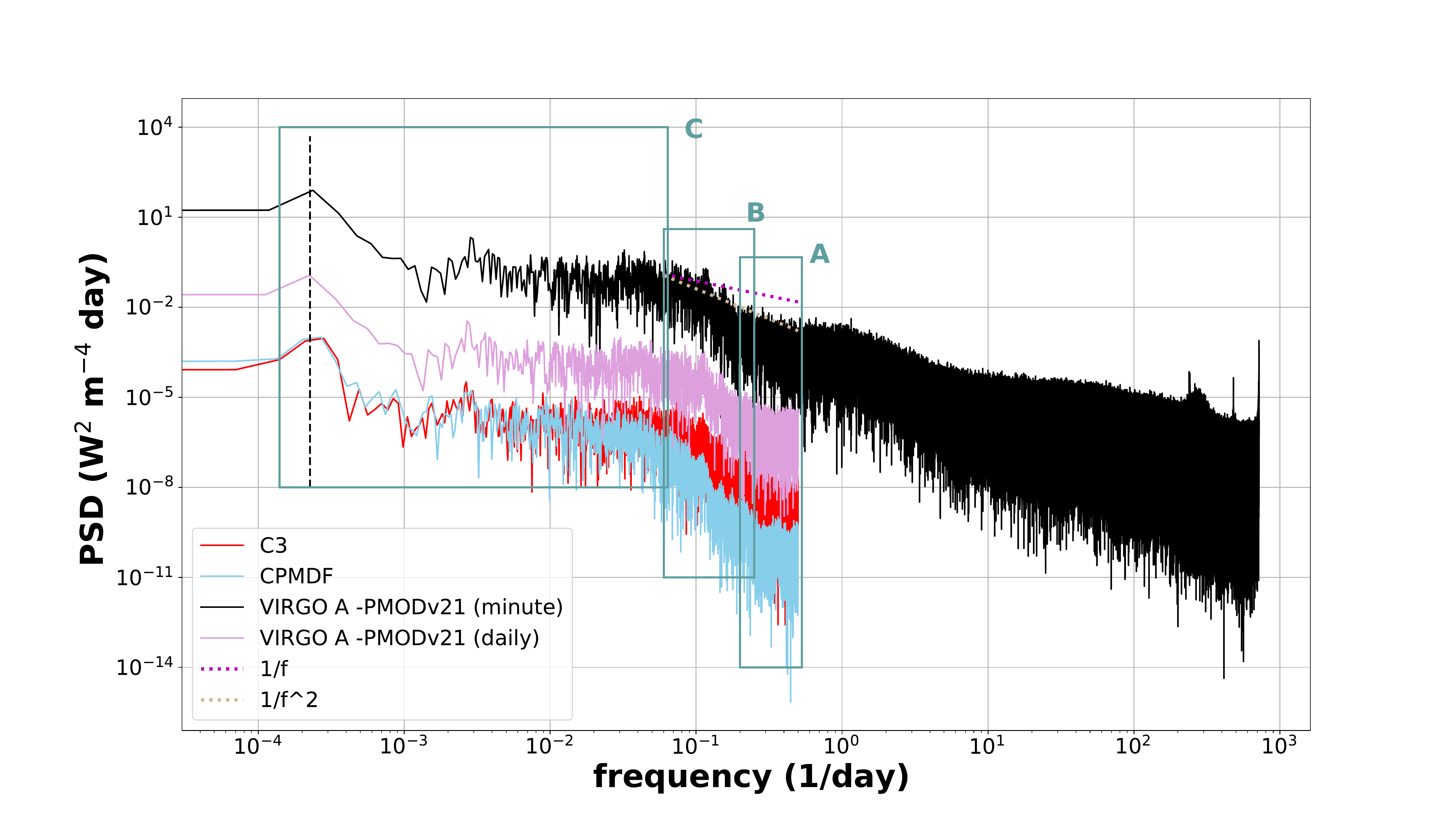}
\label{FigureA3}
\end{figure}
%

\clearpage
\section{{Assumptions and Discussion About the Data Fusion Algorithm} } \label{D-appendix}
{To further clarify the assumptions in order to fuse the various observations in Step $1$ in Section \ref{Section2.1}, we discuss in details how the Gaussian processes are applied in the specific case of the TSI data. We also define the Bayesian framework used to perform the fusion based on \citeA{Kolar2020}. The notations in Section \ref{Section2.1} are used in this appendix.} 

    {Let us define the notion of {\it Gaussian process} as used to model the solar cycle and its variations at various scales. According to \citeA{Rasmussen2006}, a Gaussian process is a generalization of the Gaussian probability distribution. A probability distribution describes random variables which are scalars or vectors (for multivariate distributions). One can loosely think of a function as a very long vector, each entry in the vector specifying the function value at a particular input (e.g., time). A Gaussian Process is a flexible distribution over functions, with many useful analytical properties. In other words, a Gaussian process is a finite linear combination of random variables with a multivariate normal distribution, completely defined by the first and second-order statistics – the mean $\mu$ and the covariance matrix - or kernel-  $k_{\mathbf{\theta}} (\mathbf{x},\mathbf{x})$. The kernel can be defined by the addition or multiplication of other kernels (periodic, linear, white, Mat\'ern … ). We assume $\mu$ to be zero, without loss of generality. The covariance function determines properties of the functions, such as smoothness, amplitude, etc.}

    {Let us model the function of interest, i.e. the solar cycle $s$ in our TSI dataset, using a GP prior, noisy observations and the associated time for each observation such as $\mathbf{s} \sim$ $GP (0, k_{\mathbf{\theta}}(t_i, t_i)_{\{i=[1,n]\}})$. $s$ is not a deterministic signal (i.e.  perfect sinusoid with an $11$-year cycle) and its variations are random (no a priori knowledge). In terms of probability distribution, we can state:}
\begin{equation*}
 p(\mathbf{s}) = \mathcal{N} (\mathbf{s}; 0, k_{\mathbf{\theta}})     
\end{equation*}
\begin{equation*}
p(\mathbf{y} | \mathbf{s}) = \Pi_{i=1}^n    \mathcal{N} (y_i; s_i, \sigma^2)
\end{equation*}
{$p(\mathbf{s})$  is the probability distribution of $s$. $p(\mathbf{y} | \mathbf{s})$ is the conditional probability of $s$ knowing $\mathbf{y} = [y_i]$. In Section \ref{Section2.1}, the observation $y_i$ is not a scalar, but a vector due to the number of input TSI time series to fuse, i.e. $\mathbf{y}_i = [a(t_i), b(t_i), c(t_i)]$ with an uncorrelated noise   $\mathbf{\sigma^2 I}$. The functions $a$, $b$ and $c$ are defined in the model of the observations in Eq. \eqref{CorrectedMeas}. $\mathbf{\sigma^2 I}$  is (assuming uncorrelated measurements) formulated as $diag([\sigma^2_a, \sigma^2_b,$  $\sigma^2_c])$. Now, the covariance matrix of $\mathbf{y}$ is defined as $k_{\mathbf{\theta}} (\mathbf{x},\mathbf{x}) + \mathbf{\sigma^2 I}$. $\mathbf{x} = [t_i, t_i]$ is the concatenation of times associated with each input time series. All the hyperparameters defining the kernel are in the vector  $\mathbf{\theta}$. The common approach is to estimate these hyperparameters to model the desired signal using a (marginal) MLE:}

\begin{equation*}
    \theta^{*} = \argmax_{\mathbf{\theta}} p(\mathbf{y} | \mathbf{\theta})
\end{equation*}

{and the estimation of $s$ ($s^{*}$) via:}

\begin{equation*}
p(\mathbf{y^*} | \mathbf{y}) = \frac{p(\mathbf{y^*} , \mathbf{y})}{\mathbf{p(y)}} = \int  p(\mathbf{y^*} | \mathbf{s^*}) p(\mathbf{s^*} | \mathbf{s}) p(\mathbf{s} | \mathbf{y}) d\mathbf{s}d\mathbf{s^*}
\end{equation*}

{Note that $*$ means the estimated parameter or signal. While the marginal likelihood, the posterior and the predictive distribution all have closed-form Gaussian expressions, the cost of evaluating them scales as $O(n^3)$ due to the inversion of $k_{\mathbf{\theta}} (\mathbf{x},\mathbf{x}) + \mathbf{\sigma^2 I}$, which is impractical for large datasets.}
{To overcome this limitation, we approximate the kernel matrix $k_{\mathbf{\theta}} (\mathbf{x},\mathbf{x})$  with a low-rank matrix $Q_{\mathbf{\theta}}$ described in Section \ref{Section2.1} as  $Q_{\mathbf{\theta}}  = k_{\mathbf{\theta}} (\mathbf{x},\mathbf{u}) k_{\mathbf{\theta}} (\mathbf{u},\mathbf{u}) ^{-1} k_{\mathbf{\theta}} (\mathbf{u},\mathbf{x})$. $\mathbf{u}$ is a vector of inducing points. It is important to underline that the vector of inducing points $\mathbf{u}$ introduced in this approximation is crucial. A proper training procedure (size of the vector of inducing points) permits us to learn about the stochastic properties of the data at various scales, which allows long-term and short-term correlations to be taken into account. Various methods exist to perform the MLE using the low-rank matrix with additional assumptions. Readers can refer to \citeA{Bauer2016} for a comprehensive description on this topic. Here, we use the maximization of the lower bound derived from the log-marginal likelihood $\log{p(\mathbf{y}|\mathbf{x})}$ based on the variational free energy method developed in \citeA{Bauer2016} and discussed in \citeA{Kolar2020}. With this approximation, one needs to consider $s$ as a sparse GP, a special class of GPs. With many data sample (i.e. $> 10^3$), the error due to the approximating the GP as a sparse GP tends to $0$ according to  \citeA{Kolar2020}.}

 {The first test of data fusion was carried out when fusing the data recorded by the radiometer PMO6V on board of the SOHO/VIRGO mission  from the two channels VIRGO/PMO6V-A and VIRGO/PMO6V-B described in \citeA{Finsterle2021}. \citeA{Kolar2020} optimzed the data fusion process using dedicated simulations. The latter were similar when more channels/observations were analyzed. We spare the patience of the Reader by describing redundant simulations.}

 {Finally, we would like to comment further on the difference of definitions between the stochastic and solar noises described in Section \ref{Section2}. In our work, solar noise is defined to cover solar activity and the small-timescale processes which produce short-term variations. The stochastic noise takes into account both the instrumental effect and the small-timescale solar activity. That together produces the random variability (i.e. white noise) in time at a small scale. The instrumental noise defines all the short-term and small amplitude solar variations which cannot be properly observed by the radiometer due to its characteristics (observation rate, distance to the event, …). Statistically, it is an uncorrelated zero-mean Gaussian noise. On a large scale, the stochastic noise can be modelled as having a power law dependency with respect to frequency, which could be related to diffusive effects. We are here speaking about the long-term correlations. It is related to the solar noise with the diffusion effects within the solar cycle, the sun's rotation period, ... etc.} 

\clearpage

\clearpage

\end{document}